\documentclass[aps,pra,twocolumn, notitlepage, superscriptaddress, nofootinbib]{revtex4-2}

\usepackage{graphicx, xcolor}
\usepackage{dcolumn}
\usepackage{bm}

\usepackage{bbm}
\usepackage{booktabs}
\usepackage[nice]{nicefrac}
\usepackage{comment}
\usepackage[linesnumbered,ruled]{algorithm2e}
\usepackage{dsfont}
\usepackage{xargs}

\usepackage{psfrag}
\usepackage{amssymb,amsmath,amsthm,enumerate,amsfonts}

\RequirePackage[colorlinks,citecolor=blue,urlcolor=blue]{hyperref}

\newtheorem{lemma}{Lemma}
\newtheorem{remark}[lemma]{Remark}
\newtheorem{proposition}[lemma]{Proposition}

\newtheorem{theorem}[lemma]{Theorem}

\newtheorem{corollary}[lemma]{Corollary}
\newtheorem*{Proposition*}{Proposition}
\newenvironment{proofsketch}{\paragraph*{Proof Sketch:}}{\hfill$\square$}

\newcommand{\II}{{\mathbb{I}}}

\newcommand{\dE}{\mathbb {E}}
\newcommand{\dP}{\mathbb{P}}
\newcommand{\dN}{\mathbb {N}}

\newcommand{\cF}{\mathcal {F}}

\newcommand{\cE}{\mathcal {E}}
\newcommand{\cA}{\mathcal {A}}
\newcommand{\cB}{\mathcal {B}}
\newcommand{\cC}{\mathcal {C}}

\newcommand{\cM}{\mathcal {M}}

\newcommand{\cN}{\mathcal {N}}

\newcommand{\cD}{\mathcal {D}}

\newcommand{\BEAS}{\begin{eqnarray*}}
\newcommand{\EEAS}{\end{eqnarray*}}
\newcommand{\BEA}{\begin{eqnarray}}
\newcommand{\EEA}{\end{eqnarray}}
\newcommand{\BEQ}{\begin{equation}}
\newcommand{\EEQ}{\end{equation}}
\newcommand{\BIT}{\begin{itemize}}
\newcommand{\EIT}{\end{itemize}}
\newcommand{\BNUM}{\begin{enumerate}}
\newcommand{\ENUM}{\end{enumerate}}

\begin{document}


\title{Performance of Gaussian Boson Sampling on Planted Bipartite Clique Detection}

\author{Yu-Zhen Janice Chen} 
\email{yuzhenchen@cs.umass.edu} 
\affiliation{University of Massachusetts Amherst}
\author{Laurent Massoulié} 
\email{laurent.massoulie@inria.fr} 
\affiliation{Inria, Paris}
\author{Don Towsley}
\email{towsley@cs.umass.edu}
\thanks{Authors are listed alphabetically.}
\affiliation{University of Massachusetts Amherst}


\date{\today}

\begin{abstract}
We investigate whether Gaussian Boson Sampling (GBS) can provide a computational advantage for solving the planted biclique problem, which is a graph problem widely believed to be classically hard when the planted structure is small. Although GBS has been heuristically and experimentally observed to favor sampling dense subgraphs, its theoretical performance on this classically hard problem remains largely unexplored.
We focus on a natural statistic derived from GBS output: the frequency with which a node appears in GBS samples, referred to as the \emph{node weight}. We rigorously analyze whether this signal is strong enough to distinguish planted biclique nodes from background nodes. Our analysis characterizes the distribution of node weights under GBS and quantifies the bias introduced by the planted structure.
The results reveal a sharp limitation: when the planted biclique size falls within the conjectured hard regime, the natural fluctuations in node weights dominate the bias signal, making detection unreliable using simple ranking strategies. These findings provide the first rigorous evidence that planted biclique detection may remain computationally hard even under GBS-based quantum computing, and they motivate further investigation into more advanced GBS-based algorithms or other quantum approaches for this problem.

\end{abstract}

\maketitle


\section{Introduction}

The \emph{planted biclique problem}~\cite{ames2011nuclear, feldman2017statistical} is a well-known graph problem conjectured to be computationally hard for classical algorithms, with a wide range of applications including molecular property identification~\cite{daminelli2012drug}, community detection in social networks~\cite{wang2015review, he2018hidden}, and financial fraud detection in transactional graphs~\cite{braun2017improving, chen2022antibenford}. 
An analogous version of this problem, the \emph{planted clique problem}~\cite{jerrum1992large}, is known to be efficiently solvable by polynomial-time algorithms when the planted clique size satisfies $K = \Omega(\sqrt{n})$~\cite{kuvcera1995expected, alon1998finding} in a random graph of size $n$ nodes/vertices, and is information-theoretically undetectable when $K \leq 2 \log n$ ~\cite{bollobas1976cliques}.  
However, in the intermediate regime $2 \log n \ll K \ll \sqrt{n}$, the computational complexity remains unresolved: no classical polynomial-time algorithm is known to succeed, and it is widely conjectured that none exists~\cite{juels2000hiding, applebaum2010public, hazan2011hard, hirahara2024planted, krivelevich2002approximating}.
Given this challenge, the planted biclique problem is expected to exhibit similar computational hardness since the planted clique problem can be reduced to it.

\emph{Gaussian Boson Sampling (GBS)}~\cite{hamilton2017gaussian, kruse2019detailed} has recently attracted increasing research interest for its potential to offer quantum computational advantages. 
While GBS and its predecessor, \emph{Boson Sampling}~\cite{aaronson2011computational}, were originally introduced to demonstrate quantum advantage in specialized sampling problems, subsequent discoveries have revealed underlying connections between GBS and graph-theoretic structures~\cite{hamilton2017gaussian, kruse2019detailed}. 
These connections have motivated a variety of proposed applications, including estimating the number of perfect matchings in undirected graphs~\cite{bradler2018gaussian}, measuring graph similarity~\cite{schuld2020measuring}, accelerating graph classification~\cite{sureka2024gaussian}, enhancing stochastic optimization algorithms~\cite{arrazola2018quantum}, identifying dense subgraphs~\cite{arrazola2018using}, and discovering high-weight cliques~\cite{banchi2020molecular}. 
Several proof-of-principle experimental demonstrations have also been reported~\cite{sempere2022experimentally, bulmer2022boundary, deng2023solving}.

These developments naturally raise the question: \emph{Can GBS offer a computational advantage for solving the classically hard planted biclique problem?}
A positive answer would suggest that GBS can significantly accelerate planted biclique or clique detection, with direct implications for practical applications such as molecular property discovery, community detection, and financial fraud detection~\cite{daminelli2012drug, wang2015review, he2018hidden, braun2017improving, chen2022antibenford}.
Conversely, a negative result would lend further support to the average-case hardness conjecture of the planted clique problem, which underpins hardness assumptions in areas such as public-key cryptography~\cite{juels2000hiding, applebaum2010public}, approximate Nash equilibria~\cite{hazan2011hard}, and independence number estimation~\cite{krivelevich2002approximating}.
In either case, addressing this question advances our understanding of both the computational capabilities of GBS and the intrinsic hardness of planted (bi)clique problems, making it a compelling direction of study.
To date, the potential of GBS for sampling dense subgraphs has been explored primarily through heuristic~\cite{arrazola2018using} and experimental~\cite{banchi2020molecular, sempere2022experimentally, deng2023solving} approaches. In this work, we aim to investigate the power of GBS in detecting planted bicliques through a rigorous analytical lens.

To investigate this question, we consider node-specific statistics as a means of distinguishing planted biclique nodes from non-planted ones in bipartite Erdős–Rényi (ER) random graphs.
To leverage the unique property of GBS -- that the sampling probability of a subgraph is proportional to the square of its number of perfect matchings (i.e., the square of its Hafnian)~\cite{arrazola2018using} -- we propose to study a statistic we refer to as the \emph{weight}.
Specifically, when GBS samples subgraphs of size $2m$, the weight of a node is defined as the sum, over all size-$2m$ bipartite subgraphs containing the node, of the squared number of perfect matchings.
This weight directly corresponds to the probability that the node appears in a GBS sample, capturing how GBS inherently favors subgraphs with rich combinatorial structure. 
While it is not a priori clear whether this statistic can distinguish planted biclique nodes from non-planted biclique nodes, it provides a natural and GBS-informed lens to explore the planted biclique problem.

Our analysis of GBS-informed vertex weights in the planted biclique setting yields the following key insights:
\begin{enumerate}
    \item 
    We prove that the joint moments of centered and rescaled vertex weights converge in distribution to a multivariate Gaussian, with an explicit covariance structure derived via probabilistic and combinatorial analysis. 
    This result provides a rigorous statistical foundation for understanding the fluctuation and correlation of GBS sampling frequencies across nodes.

    \item 
    We derive precise asymptotics for the expected weights of planted biclique nodes and background nodes. We found that planted biclique nodes have higher weights than background nodes. Hence, GBS sampling \emph{bias} toward sampling planted biclique nodes, with the bias scaling as a \(K/n\) fraction of the mean weight, providing a quantitative measure of the planted biclique signal strength.

    \item 
    Combining the bias and variance estimates, we show that in the regime \(K = o(\sqrt{n})\), the bias becomes negligible relative to the standard deviation, indicating limited statistical power to reliably distinguish planted biclique nodes in this regime. Though, when \(K = \Theta(\sqrt{n})\), the bias becomes detectable, and planted biclique nodes can be statistically identified with high probability.
\end{enumerate}

This paper is organized as follows:
Section~\ref{sec:formulation} introduces the background and formalizes the planted biclique problem, along with the application of GBS to subgraph sampling.
Section~\ref{sec:method} presents the key GBS-derived statistic, referred to as the \emph{weight} of a node, and outlines the key probabilistic tools used throughout our analysis.
Section~\ref{sec:analysis-planted} analyzes the performance of GBS in detecting planted bicliques by characterizing joint moments of node weights, quantifying the weight bias favoring planted biclique nodes, and examining their implications for the detection task.
Section~\ref{sec:discussion} discusses broader implications, potential future directions, and highlights a subsidiary result on characterizing Hafnians in bipartite Erdős–Rényi graphs.
We conclude in Section~\ref{sec:conclu}, with technical proofs deferred to the appendices.

\section{Background and Problem Setup}\label{sec:formulation}

In this section, we first formalize the planted biclique problem~\cite{ames2011nuclear, feldman2017statistical} and illustrate the limitations of simple statistical approaches, such as degree-based statistics, in distinguishing planted biclique nodes from non-planted/background nodes when the biclique size is $o(\sqrt{n})$. 
We then introduce Gaussian Boson Sampling (GBS)~\cite{hamilton2017gaussian, kruse2019detailed}  as a subgraph sampling framework whose output distribution favors subgraphs with many perfect matchings~\cite{arrazola2018using}.

\subsection{Planted Biclique Problem}\label{sec:planted-biclique-problem}

In the {planted biclique problem}~\cite{ames2011nuclear, feldman2017statistical}, one is given a graph $\hat{G}$ that is equally likely to be either a bipartite Erdős–Rényi (ER) graph \( \hat{G} = G \) or the same graph with a planted biclique \( \hat{G} = G' \). The objective is to determine, with high probability, which of the two graphs was given, or, in the case where \( \hat{G} = G' \), to locate the planted biclique subgraph.

Specifically, we denote \( G \sim \mathrm{ER}(n, n, p) \) as a bipartite ER graph with vertex set \( V \cup V' \), where 
\( |V| = |V'| = n \). The edge set \( E \subseteq V \times V' \) is generated by including each possible edge independently with some fixed probability \( p \in (0,1) \). Specifically, for each \( u \in V \) and \( v \in V' \), let the indicator variable
\begin{align}
    \xi_{uv} := \II_{\{\text{edge } (u,v) \text{ is present in } G\}}
\end{align}
denote whether the edge \( (u,v) \) exists. We assume that variables \(\xi_{uv}\) are mutually independent Bernoulli random variables with parameter \( p \), i.e., \( \xi_{uv} \sim \mathrm{Ber}(p) \).

We then let \( G' \sim \mathrm{ER}(n, n, p, K, K) \) denote a bipartite ER graph with a planted biclique of size \( K \times K \), where \( K = \epsilon \sqrt{n} \) {for some \( \epsilon > 0 \).} The graph \( G' \) is constructed by first generating \( G \sim \mathrm{ER}(n, n, p) \) and then selecting uniformly at random vertex subsets \( \mathcal{A}_0 \subset V \) and \( \mathcal{B}_0 \subset V' \) with \( |\mathcal{A}_0| = |\mathcal{B}_0| = K \). We then plant the biclique by setting
\begin{align}
    \xi_{uv} = 1 \quad \text{for all } u \in \mathcal{A}_0, \; v \in \mathcal{B}_0,
\end{align}
ensuring that all possible edges in \( \mathcal{A}_0 \times \mathcal{B}_0 \) are present in \( G' \), regardless of whether they were present in \( G \). Note that in this paper, we refer to nodes in $\cA_0 \cup \cB_0$ as planted nodes, and refer to $\{V \cup V'\} \setminus \{\cA_0 \cup \cB_0\}$ as non-planted/background nodes.

The planted biclique problem shares the difficulty of the planted clique problem~\cite{jerrum1992large, kuvcera1995expected, alon1998finding}, where the objective is to differentiate a random graph from one containing a planted clique.
The planted clique problem can be reduced to the planted biclique problem by constructing a bipartite graph with two vertex sets: one corresponding to the original graph's vertices, and the other consisting an identical copy. Edges are then added between these two sets of nodes based on the adjacency structure of the original graph, and each vertex is also connected to its corresponding copy. Under this transformation, detecting a planted biclique in the resulting bipartite graph is equivalent to detecting a planted clique in the original graph. Consequently, the planted biclique problem is at least as hard as the planted clique problem.

When $2\log n \ll K \ll \sqrt{n}$, the best algorithm known for the planted biclique problem remains to be a brute-force search~\cite{chen2019broadcast}. 
This method systematically examines all balanced bipartite subgraphs of size slightly larger than what could occur naturally, e.g., \( 3\log n \times 3\log n\) (which moderately exceeds the information theoretic threshold \( 2\log n \)), and checks whether any of them form a biclique. 
The discovery of such a biclique would signal the presence of a planted structure.
This approach operates in quasi-polynomial time.
To date, no proof exists ruling out a polynomial-time algorithm for detecting planted cliques in the regime $2\log n \ll K \ll \sqrt{n}$, though some conjecture that no such classical algorithm exists~\cite{juels2000hiding, applebaum2010public, hazan2011hard, hirahara2024planted, krivelevich2002approximating}.

To further illustrate the difficulty of the planted biclique problem in the \emph{conjectured hard regime} \( 2\log n \ll K \ll \sqrt{n} \), we analyze the distribution of vertex degrees and show that even this natural statistic fails to distinguish a planted node \( i \in \mathcal{A}_0 \) from a non-planted node \( i' \notin \mathcal{A}_0 \).
Let \( D(i) \) denote the degree of node \( i \), i.e., the number of its neighbors in the bipartite graph. 
Then, for a {non-planted} node \( i' \notin \mathcal{A}_0 \), its neighbors are formed by independent edge sampling with probability \( p \), so
\begin{align}
    D(i') \sim \mathrm{Bin}(n, p) \approx \mathcal{N}(np, np(1-p)),
\end{align}
where the approximation holds by the Central Limit Theorem.
A {planted biclique} node \( i \in \mathcal{A}_0 \) has exactly \( K \) deterministic connections to the planted biclique subset \( \mathcal{B}_0 \), and the remaining \( n-K \) potential neighbors are sampled with probability \( p \). Thus,
\begin{align}
    D(i) &\sim K + \mathrm{Bin}(n - K, p) \\
    &\approx K + \mathcal{N}((n-K)p, (n-K)p(1-p)).
\end{align}

To compare the two degrees, we center and normalize both and apply approximations using the fact that $K = o(\sqrt{n})$:
\begin{align}
    &\tilde{D}(i') := \frac{D(i) - np}{\sqrt{np(1-p)}}\approx \mathcal{N}(0,1), \\
    &\tilde{D}(i) := \frac{D(i') - np}{\sqrt{np(1-p)}}\approx \frac{K(1 - p)}{\sqrt{np(1 - p)}} + \mathcal{N}(0,1).
\end{align}
The key difference lies in the \emph{mean shift} \( \frac{K(1 - p)}{\sqrt{np(1 - p)}} \) 
which represents the signal distinguishing a planted biclique node from a non-planted node based on degree.
However, as \( K = o(\sqrt{n}) \), 
\begin{align}
    \frac{K(1 - p)}{\sqrt{np(1 - p)}} \rightarrow 0 \text{ as } n \rightarrow \infty,
\end{align}
meaning the mean shift vanishes and the two normalized distributions become indistinguishable in total variation. 
Hence, degree-based heuristics cannot reliably detect planted biclique nodes when \( K \ll \sqrt{n} \), reinforcing the intuition that the planted biclique detection problem is computationally and statistically hard in this regime.

\subsection{Gaussian Boson Sampling}\label{sec:applying-GBS}
Gaussian Boson Sampling (GBS)~\cite{hamilton2017gaussian, kruse2019detailed} is a special-purpose linear optical quantum computing paradigm that involves preparing Gaussian states as inputs, passing them through a Haar-random interferometer, and measuring the output states in the Fock basis.
Consider an idealized GBS system with $2n$ modes, which we use to sample subgraphs from a balanced bipartite graph with \( 2n \) nodes in total. We denote the output pattern of GBS by ${\bf s} = (s_1, s_2, ..., s_{2n})$, where $s_i$ is the number of photons detected in output mode $i$. 
Let $2m = s_1 + s_2 + ... + s_{2n}$ be the number of photons prepared in the input state.

We can control the input photon number \(2m\) by selecting an appropriate number of Single Mode Squeezed States (SMSS) and their squeezing parameters. According to \cite[Eq. (12)]{hamilton2017gaussian} \cite[Eq. (32)]{kruse2019detailed}\cite{weedbrook2012gaussian}, by adjusting the number of input SMSSs, denoted by \( a \), and the squeezing parameter \( b \), we can achieve \( 2m = (a - 1) \text{sinh}^2(b) \) photons in the input with the following probability:
\begin{align}
    \mathbb{P}(2m|a, b) = {\frac{a}{2}+2m-1 \choose 2m} \text{sech}^{a}(b) \text{sinh}^{2m}(b).
\end{align}
We follow the convention in the literature~\cite[Section 6.2]{aaronson2011computational}\cite{hamilton2017gaussian, kruse2019detailed} restricting $m = \epsilon \sqrt{2n}$ {for some fixed constant 
$\epsilon \in (0, 1)$} so that $2m < \sqrt{2n}$. This limitation ensures that with high probability the output pattern remains collision-free, i.e., $s_i \in \{0, 1\}, \forall i \in \{1,..., 2n\}$, according to the Bosonic birthday paradox~\cite[Appendix 13]{aaronson2011computational} \cite{arkhipov2012bosonic}. 

Hamilton et al.~\cite{hamilton2017gaussian} and Kruse et al.~\cite{kruse2019detailed} show that the probability of observing an output pattern ${\bf s}$ is 
\begin{align}
    \mathbb{P}({\bf s}) = \frac{\text{Haf}(\mathcal{M}_{\bf s})}{s_1!s_2!\cdots s_{2n}! \sqrt{|\sigma_Q|}},
\end{align}
where the function $\text{Haf}(\mathcal{M}_{\bf s})$ is the Hafnian of matrix $\mathcal{M}_{\bf s}$, matrix $\mathcal{M}_{\bf s}$ is a submatrix of matrix $\mathcal{M}$ fixed by ${\bf s}$ (see~\cite[Figure 2]{kruse2019detailed}), 
\begin{align}\label{eq:encode-1}
    &\mathcal{M} = 
    \begin{pmatrix}
        0 & \mathbbm{1}_{2n} \\
        \mathbbm{1}_{2n} & 0
    \end{pmatrix} 
    \left[\mathbbm{1}_{4n}-\sigma_Q^{-1}\right],
    &\sigma_Q = \sigma + \frac{1}{2}\mathbbm{1}_{4n}
\end{align}
are matrices characterized by the $(4n \times 4n)$-dimensional covariance matrix $\sigma$ of the $2n$-mode Gaussian state, and $|\sigma_Q|$ denotes the determinant of matrix $\sigma_Q$. 
Brádler et al.~\cite{bradler2018gaussian} further show that one can encode a graph adjacency matrix $M$ (of size $2n \times 2n$) into a Gaussian state by letting
\begin{align}\label{eq:encode-2}
    \mathcal{M} = c
\begin{pmatrix}
    M & 0\\
    0 & M
\end{pmatrix},
\end{align}
where $c < \lambda^{-1}$ is a constant, and $\lambda$ is the largest eigenvalue of $M$. 
Then, combining~\eqref{eq:encode-1} and~\eqref{eq:encode-2}, one can derive the corresponding covariance matrix \(\sigma\) from the adjacency matrix \(M\), and use it to prepare the appropriate input Gaussian state.

We will also post-select output samples from GBS and only keep samples such that $s_i \in \{0, 1\}$ and $\sum_{i}^{2n} s_i= 2m$. Then, according to Arrazola and Bromley~\cite{arrazola2018using}, the probability of getting such kind of output pattern ${\bf s}$ is
\begin{align}\label{eq:GBS-prob}
    \mathbb{P}_{k\text{cf}}({\bf s}) = \frac{c^2|\text{Haf}(M_{\bf s})|^2}{\sqrt{|\sigma_Q|}\mathbb{P}(2m)} \propto |\text{Haf}(M_{\bf s})|^2,
\end{align}
where $M_{\bf s}$ is the adjacency matrix corresponding the subgraph of $M$ selected by ${\bf s}$, i.e., treating $(s_1, s_2, ..., s_{2n})$ as indicators of selecting the corresponding vertex or not.
Note that the Hafnian of an adjacency matrix is equivalent to the number of perfect matchings in the corresponding graph, i.e., the number of matchings that cover every vertex in the graph. 
Hence,~\eqref{eq:GBS-prob} essentially implies that the more perfect matchings a subgraph has, the more likely it is sampled by GBS.

\section{Methodology}\label{sec:method}

In this section, we introduce the key definitions, concepts, and lemmas that form the foundation of our analysis. 
We begin by formally defining vertex/node \emph{weights} in terms of the probability that a node appears in an idealized Gaussian Boson Sampling (GBS) output sample, which essentially reduces to computing the expected number of perfect matchings in the corresponding GBS-induced subgraphs. We then describe how these weights can be used to design a GBS-based algorithm for planted biclique detection, leveraging the frequency of vertex occurrences in the sampled subgraphs. Following this, we present two essential probabilistic tools: a Poisson approximation for the intersection sizes of randomly selected vertex subsets, and a characterization of the probability that two random bijections agree on overlapping domains.
Together, these results enable us to derive rigorous characterizations and approximations for the weight distribution and for the weight differences between planted and non-planted nodes.

\subsection{Definitions of Weights}\label{sec:def-weight}

According to \eqref{eq:GBS-prob}, the probability that idealized GBS outputs a specific subset $\mathcal{S}$ of $2m$ vertices is proportional to the square of the number of perfect matchings in the subgraph induced by $\mathcal{S}$. When the input graph $\hat{G} = (V \cup V', E)$ is bipartite, the sampled subset $\mathcal{S}$ must take the form of a pair $(\cA, \cB)$, where $\cA \subset V$, $\cB \subset V'$, and $|\cA| = |\cB| = m$. This restriction arises because an unbalanced bipartite subgraph cannot contain a perfect matching and is therefore excluded from the sampling process.
For a given pair $(\cA, \cB)$, we define the normalized expected number of perfect matchings in $(\cA, \cB)$ as
\begin{align}
    Y(\cA, \cB) := \frac{1}{p^m m!} \sum_{\sigma \in \mathrm{Bij}(\cA, \cB)} \prod_{u \in \cA} \xi_{u\sigma(u)},
\end{align}
where $\mathrm{Bij}(\cA, \cB)$ denotes the set of all bijections from $\cA$ to $\cB$ (i.e., $\sigma \in \mathrm{Bij}(\cA, \cB)$, $\sigma: \cA \rightarrow \cB$), $\xi_{uv}$ indicates the presence of edge $(u, v)$, and $p$ is the probability that each edge exists. The probability of observing $(\cA,  \cB)$ in the GBS output is then proportional to $Y(\cA, \cB)^2$.

To quantify the contribution of a vertex $i \in V = [n]$ to the GBS output, we define its \emph{weight} as the normalized summation of $Y(\cA, \cB)^2$ over all possible $(\cA,  \cB) \in (V, V')$ that contains node $i$, i.e., 
\begin{align}
W(i) &:= \frac{1}{\binom{n-1}{m-1} \binom{n}{m}} \sum_{\substack{\mathcal{A} \in \binom{V}{m} \\ i \in \mathcal{A}}} \sum_{\mathcal{B} \in \binom{V'}{m}} Y(\mathcal{A}, \mathcal{B})^2 \\
&= \frac{1}{\binom{n-1}{m-1} \binom{n}{m}} \sum_{\substack{\mathcal{A} \in \binom{V}{m} \\ i \in \mathcal{A}}} \sum_{\mathcal{B} \in \binom{V'}{m}} \bigg( \frac{1}{p^{2m}(m!)^2} \times\notag \\
&\qquad\qquad\qquad \sum_{\substack{\sigma, \tau \in \mathrm{Bij}(\mathcal{A}, \mathcal{B})}} \prod_{u \in \mathcal{A}} \xi_{u\sigma(u)} \xi_{u\tau(u)} \bigg), \label{eq:W_as_sum}
\end{align}
where $\binom{V}{m}$ denotes the set of all size-$m$ subsets of $V$.

Alternatively, $W(i)$ can be expressed as an expectation over random subsets and bijections:
\begin{align}
W(i) \equiv \mathbb{E}_{\cA, \cB, \sigma, \tau} \left[ \prod_{u \in \mathcal{A}} \frac{\xi_{u\sigma(u)} \xi_{u\tau(u)}}{p^2}  \right], \label{eq:W}
\end{align}
where the expectation is taken over a uniformly random subset $\mathcal{A} \subset V$ of size $m$ containing $i$, a uniformly random subset $\mathcal{B} \subset V'$ of size $m$, and independent bijections $\sigma, \tau \in \mathrm{Bij}(\mathcal{A}, \mathcal{B})$, conditioned on the edge set $E$ of the input graph $\hat{G}$. Weights for nodes $i \in V'$ are defined analogously.

The probability that GBS outputs a vertex set containing node $i$ is proportional to its weight $W(i)$. When the input graph $\hat{G}$ contains a planted biclique with vertex sets $\mathcal{A}_0 \subset V$ and $\mathcal{B}_0 \subset V'$, the expected weight of node $i$ can be expressed as
\begin{align}
\mathbb{E}[W(i)] = \mathbb{E} \left[ p^{-S_1} \cdot p^{-S_2} \right], \label{eq:W_expectation}
\end{align}
where the expectation is taken over all randomness, including the randomness of the input graph; the random variable
\begin{align}
    S_1 = \sum_{u \in \mathcal{A} \cap \mathcal{A}_0} \left( \mathbb{I}_{\sigma(u) \in \mathcal{B}_0} + \mathbb{I}_{\tau(u) \in \mathcal{B}_0} \right)
\end{align}
counts the number of times the bijections/matchings intersect the planted biclique, and
\begin{align}
    S_2 = \sum_{u \in \mathcal{A}} \mathbb{I}_{\sigma(u) = \tau(u) \,\&\, (u, \sigma(u)) \notin \mathcal{A}_0 \times \mathcal{B}_0}
\end{align}
counts the number of collisions between the two matchings that occur outside the planted biclique structure.
 
Given the definition of vertex weights above, it is natural to consider a GBS-based approach for planted biclique detection, as outlined in Algorithm~\ref{alg:GBSClique}. The algorithm repeatedly samples vertex subsets from the given graph using idealized GBS. For each vertex $i$, it records the fraction of sampled subsets in which $i$ appears. This empirical frequency is proportional to the vertex's weight $W(i)$, as defined earlier. The decision function \( f \) can be instantiated with any statistic based on the vertex weights; we keep this choice general for now and will examine a specific instantiation in Section~\ref{sec:planted-consequences}.

\begin{algorithm}[tp]
    \caption{GBS-Based Planted Biclique Detection}\label{alg:GBSClique}
        \textbf{Input:} graph $\hat{G}$, decision function $f$, horizon $T$
        
        \textbf{Initialize:} prepare GBS sampler $\texttt{GBS}_{\hat{G}}(2m)$; set $\hat{W}(i) \gets 0$ for all $i \in V \cup V'$

        \For{$t = 1$ to $T$}{
            $\mathcal{S}_t \gets \texttt{GBS}_{\hat{G}}(2m)$\;
            
            \For{each $i \in \mathcal{S}_t$}{
                $\hat{W}(i) \gets \hat{W}(i) + 1$ \tcp*{Counting occurrences}
            }
        }
        \For{each $i \in V \cup V'$}{
            $\hat{Z}(i) \gets \frac{\hat{W}(i)}{T}-\frac{2m}{2n}$ \tcp*{Centering}
        }
        $\sigma^2 \gets \frac{1}{2n} \sum_{i \in V\cup V'} \left(\hat{Z}(i)\right)^2$\;   
        
        \For{each $i \in V \cup V'$}{
            $\hat{Z}(i) \gets \frac{\hat{Z}(i)}{\sigma}$ \tcp*{Normalizing}
        }
        \uIf{$f((\hat{Z}(i))_{i \in V\cup V'})$}{
            \textbf{Output:} \texttt{True} \tcp*{Planted biclique detected}
        }
        \Else{
            \textbf{Output:} \texttt{False} \tcp*{No planted biclique}
        }
\end{algorithm}

\subsection{Key Technical Tools}\label{sec:key-tools}

Having introduced the definition of vertex weight, we now present two key lemmas that serve as foundational tools for our subsequent analysis. These results characterize the behavior of intersection sizes among random vertex subsets and the alignment patterns of random bijections, both of which play a central role in understanding the combinatorics of GBS sampling.
We begin with Lemma~\ref{lem:poisson_intersections}, which approximates the distribution of intersection sizes among independently sampled vertex subsets. This is followed by Lemma~\ref{lem:stein_chen_permutations}, which describes the distribution of number of agreements between two independent bijections defined over highly overlapping vertex subsets. In both cases, we show that the relevant quantities behave like Poisson random variables with small mean parameters.

To formalize the first lemma, let $\mathcal{B}_0, \ldots, \mathcal{B}_\ell$ (with fixed $\ell = O(1)$) denote vertex subsets independently sampled from a ground set $V'$ of $n$ vertices, where each $\mathcal{B}_j$ is drawn uniformly at random from $\binom{V'}{m_j}$ {with $m_j = \Theta(\sqrt{n})$.} For every subcollection of vertex subsets $\mathcal{C} \subseteq \{0,\ldots,\ell\}$, define
\begin{align}
    b_{\mathcal{C}} := \left| \left( \bigcap_{j \in \mathcal{C}} \mathcal{B}_j \right) \big\backslash \left( \bigcup_{j \notin \mathcal{C}} \mathcal{B}_j \right) \right|,
\end{align}
which counts the number of elements in $V'$ that belong to exactly those subsets $\mathcal{B}_j$ for $j \in \mathcal{C}$.

We characterize the distribution of $b_{\mathcal{C}}$ for all $\mathcal{C}$ with $|\mathcal{C}| \ge 2$ as following:
\begin{lemma}[Vertex Subsets Intersection Size]
\label{lem:poisson_intersections}
Let $\mathcal{C} \subseteq \{0,\ldots,\ell\}$ with $|\mathcal{C}| \ge 2$, and define
\begin{align}
    \mu_{\mathcal{C}} := \frac{\prod_{j \in \mathcal{C}} m_j}{n^{|\mathcal{C}| - 1}}  = O\left(n^{-\left(\frac{|\cC|}{2}-1\right)}\right) = O(1).
\end{align}
Fix integer values $x_{\mathcal{C}} \in \mathbb{N}$ for all $\mathcal{C}$ with $|\mathcal{C}| \ge 2$, and let
\begin{align}
    x_{\mathrm{tot}} := \sum_{\mathcal{C} : |\mathcal{C}| \ge 2} x_{\mathcal{C}}.
\end{align}
Then, the following holds:
\begin{itemize}
    \item If $x_{\mathrm{tot}} \le n^{1/4}$, then
    \begin{align}
        \mathbb{P}(b_{\mathcal{C}} = x_{\mathcal{C}}, \forall |\mathcal{C}| \ge 2)
    = \left(1 + O\left(\frac{x_{\mathrm{tot}}^2}{\sqrt{n}}\right)\right)
    \prod_{|\mathcal{C}| \ge 2} e^{-\mu_{\mathcal{C}}} \frac{\mu_{\mathcal{C}}^{x_{\mathcal{C}}}}{x_{\mathcal{C}}!}.
    \end{align}
    \item If $x_{\mathrm{tot}} > n^{1/4}$, then
    \begin{align}
        \mathbb{P}(b_{\mathcal{C}} = x_{\mathcal{C}}, \forall |\mathcal{C}| \ge 2)
    \le \left(1 + O\left(\frac{x_{\mathrm{tot}}^2}{\sqrt{n}}\right)\right)
    \prod_{|\mathcal{C}| \ge 2} e^{-\mu_{\mathcal{C}}} \frac{\mu_{\mathcal{C}}^{x_{\mathcal{C}}}}{x_{\mathcal{C}}!}.
    \end{align}
\end{itemize}
\end{lemma}
Lemma~\ref{lem:poisson_intersections} shows that the sizes of specific intersections of random subsets, $b_{\cC}$, can be well approximated (or bounded) by independent Poisson distributions with small means $\mu_{\mathcal{C}} = O(1)$. We defer the proof of Lemma~\ref{lem:poisson_intersections} to Appendix~\ref{sec:proof-poisson_intersections}. 

We next analyze the structure of intersecting bijections, conditioned on significant overlap between the involved subsets:
\begin{lemma}[Bijections Intersection Size]\label{lem:stein_chen_permutations}
Let vertex sets $\cA, \cB, \cA', \cB'$ satisfy $|\cA|=|\cB|$, $|\cA'|=|\cB'|$, and  
\begin{align}
    |\cA|,\,\,|\cB|,\,\, |\cA'|,\,\,|\cB'|,\,\, |\cA\cap \cA'|,\,\, |\cB\cap \cB'|=m-o(m).
\end{align}
Let $\sigma$ and $\tau$ be independent bijections sampled uniformly at random from $\hbox{Bij}(\cA,\cB)$ and $\hbox{Bij}(\cA',\cB')$, respectively. Then, the following holds:
{
\begin{align}\label{eq:lemma_2_eq1}
    &\mathbb{P}\left(\sum_{i\in \cA\cap \cA'}\II_{\sigma(i)=\tau(i)}=0\right)=e^{-1}+o(1),\\
    \label{eq:lemma_2_eq2}
    &\mathbb{E}\left[p^{-\sum_{i\in \cA\cap \cA'}\II_{\sigma(i)=\tau(i)}}\right] = e^{p^{-1}-1} + o(1).
\end{align}
}
\end{lemma}
Lemma~\ref{lem:stein_chen_permutations} shows that the number of edges for which two independent bijections agree behaves approximately like a Poisson random variable with mean $1+o(1)$, assuming sufficient overlap between the domain and codomain sets. 
We defer the proof of Lemma~\ref{lem:stein_chen_permutations} to Appendix~\ref{sec:proof-stein_chen_permutations}.

Together, Lemmas~\ref{lem:poisson_intersections} and~\ref{lem:stein_chen_permutations} enable precise probabilistic reasoning about the edge and matching structures that emerge in the analysis of GBS weights, serving as core building blocks for the main results presented in the following section.

\section{Characterization of Node weight}\label{sec:analysis-planted}

In this section, we analyze the statistical behavior of the vertex weights as defined in \eqref{eq:W_as_sum}. 
We first characterize the joint moments of the centered and rescaled weights.
By enumerating edge configurations from random matchings, we show that these moments converge to those of a multivariate Gaussian distribution using Wick’s formula.
We then quantify the weight difference (bias) between planted and non-planted nodes, and derive precise asymptotic expressions for the mean and covariance structure.
These results reveal that when the planted biclique size falls in $K = o(\sqrt{n})$, the bias is asymptotically negligible compared to the standard deviation, making it statistically difficult to distinguish planted nodes from non-planted ones based on GBS sampling frequencies; though, when $K = \Theta(\sqrt{n})$, the signal becomes statistically detectable with high probability.

\subsection{Evaluation of Moments}\label{sec:planted-moments}

To better highlight the random fluctuations of the vertex weights and to simplify the analysis of their covariance and asymptotic behavior, we consider the \emph{centered} and \emph{scaled} version of the weight of vertex \(i\), defined as its deviation from the expected value with a \(\sqrt{n}\) scaling reflecting their fluctuation magnitude:
\begin{align}
    Z(i) := \sqrt{n}\left(W(i) - \mathbb{E}[W(i)]\right).
\end{align}
Our goal is to study the joint behavior of these centered and scaled weights in the large-graph limit \(n \rightarrow \infty\). To this end, we define the \(\ell\)-th order joint moment:
\begin{align}\label{eq:initial_def_phi}
    \phi_\ell(i(1), \ldots, i(\ell)) := \mathbb{E}\left[\prod_{j=1}^\ell Z(i(j))\right],
\end{align}
which captures correlations among \(\ell\) centered weights.

{Here, the indices \(i(1), \ldots, i(\ell)\) specify which vertex weights are involved in the joint moment.}
The indexing is constructed as follows: we fix integers \(\ell > 0\) and \(0 \le \ell_0 \le \ell\), and let \(i'(1), \ldots, i'(\ell) \in [1,..., K]\) fixed and be a collection of indices (possibly with repetitions).
We then define the index mapping by setting
\begin{align}
    i(j) = \begin{cases}
i'(j) & \text{if } j \le \ell_0, \\
i'(j) + K & \text{if } j > \ell_0,
\end{cases}
\end{align}
so that the first \(\ell_0\) terms refer to vertices in the planted biclique node set \(\mathcal{A}_0\), which we assume without loss of generality to be \([1,...,K]\), while the remaining terms refer to vertices outside of \(\mathcal{A}_0\).
For example, let \( \ell = 6 \), \( \ell_0 = 3 \), and suppose \( K = 10 \). 
Consider the base indices \(( i'(1), \ldots, i'(5) )= (1, 3, 1, 2, 1, 7) \). 
Then, according to the indexing rule, we construct:
\begin{align}
    (i(1), i(2), i(3), i(4), i(5), i(6)) = (1, 3, 1, 12, 11, 17).
\end{align}
In this example, the first three indices \(i(1), i(2), i(3)\) refer to vertices in the planted biclique \(\mathcal{A}_0 = [1, \ldots, 10]\), while the remaining three indices \(i(4), i(5), i(6)\) refer to vertices outside of \(\mathcal{A}_0\), shifted by \(K = 10\).

Hence, \(\phi_\ell(i(1), \ldots, i(\ell))\) represents the \(\ell\)-th joint moment of the centered and scaled weights for a specified configuration of vertices, and serves as a central object in understanding the statistical structure of the GBS output in the presence of a planted biclique. We characterize \(\phi_\ell(i(1), \ldots, i(\ell))\) as following:
\begin{theorem}[Joint Moment of Centered and Rescaled Weights]\label{thm:moment}
The quantity $\phi_\ell(i(1),\ldots,i(\ell))$ defined in \eqref{eq:initial_def_phi} verifies
\begin{equation}\label{eq:wick}
\phi_\ell(i(1),\ldots,i(\ell))= o(1)+ \sum_{\mu \in P_{\ell}^2} \prod_{\{j,j'\}\in \mu}C_{i(j),i(j')}
\end{equation}
where $P_{\ell}^2$ denotes the set of all partitions of the set $\{1, ..., \ell\}$ into subsets of size $2$ and
\begin{equation}\label{eq:covariances}
C_{i(j),i(j')}:=4(p^{-1}-1)e^{2(p^{-1}-1)}\left[\II_{i(j)=i(j')}+\frac{m^2}{n}\right].
\end{equation}
\end{theorem}
The full proof of Theorem~\ref{thm:moment} is presented in Appendix~\ref{sec:proof-thm-moment}. In the following, we give a high-level proof sketch that summarizes the key ideas and techniques. 
\begin{proofsketch}
To analyze the joint moment expression $\phi(i(1), \ldots, i(\ell))$, we first reformulate it using a structured representation of edge sets induced by random bijections between vertex subsets.  
A crucial observation is that most bijection configurations contribute negligibly to the joint moment unless every matching shares at least one edge with another. 
This motivates a decomposition of the expression into dominant and negligible parts.
We isolate the dominant contribution by focusing on the case where the set of shared edges for some index $j$ contains exactly one edge and these edges form a perfect matching on $[1, ..., \ell]$. 
The negligible terms, where the set of shared edges for some index $j$ contains more than one edge or where shared edges do not form a perfect matching, are shown to be small using probabilistic tail bounds and moment estimates. 
The main term is then evaluated using the two lemmas introduced in Section~\ref{sec:method}, derangement asymptotics, and Poisson approximation for the number of fixed points in a random permutation. 
The argument combines combinatorial decomposition, probabilistic bounds, conditional constructions via controlled edge switching, and classical moment methods.
\end{proofsketch}

\begin{corollary}[Weak Convergence to Multivariate Gaussian]
The collection of centered and rescaled random variables $\{Z_{i(j)}\}_{j\in [\ell]}$ has asymptotically the same moments of any order as a centered mutlivariate Gaussian vector with covariance matrix $(C_{i(j),i(j')})_{j,j'\in [\ell]}$. As a consequence, it converges weakly to this multivariate Gaussian distribution as $n\to\infty$.
\end{corollary}
\begin{proof}
The final expression \eqref{eq:wick} for $\phi(i(1),\ldots,i(\ell))$ is precisely Wick's formula (a.k.a, Isserlis's theorem~\cite{isserlis1918formula}) for the expectation of the product $X_{i(1)}\cdots X_{i(\ell)}$  of centered Gaussian random variables $X_{i(j)}$ with covariances $C_{i(j),i(j')}$.
\end{proof}

\begin{remark}
    Note that in the expression of $C_{i(j), i(j')}$ in~\eqref{eq:covariances}, if $i(j) \neq i(j')$, the indicator term is $0$, and only the $m^2/n$ term remains, reflecting weak correlations between nodes when $m < \sqrt{n}$.
\end{remark}

\subsection{Evaluation of Bias}\label{sec:planted-bias}

A planted biclique vertex is connected to more edges and participates in more perfect matchings, and thus is expected to have a larger weight intuitively. In what follows, we quantify this bias by evaluating the expected difference in weights between a planted biclique vertex \( i \in \mathcal{A}_0 \) and a non-planted/background  vertex \( i' \in V \setminus \mathcal{A}_0 \). Specifically, we analyze the quantity \( \mathbb{E}[W(i) - W(i')] \).

\begin{proposition}\label{prop:bias}
{ For $K=\Theta(\sqrt{n})$, $m=\Theta(\sqrt{n})$,} the expectation of $W(i)$ verifies
\begin{equation}\label{eq:asymp_expected_weight}
\dE [W(i)]=e^{p^{-1}-1}+o(1).
\end{equation}
Moreover, for two distinct vertices $i\in \cA_0$, $i'\notin \cA_0$, the bias $\dE [W(i)] -\dE [W({i'})]$ verifies
\begin{equation}\label{eq:asymp_bias}
\dE [W(i)] -\dE [W({i'})]=(1+o(1))e^{p^{-1}-1}(p^{-1}-1)\frac{2K}{n}.
\end{equation}
\end{proposition}
The full proof of Proposition~\ref{prop:bias} is given in Appendix~\ref{sec:proof-prop-bias}.
\begin{proofsketch}
We aim to evaluate the expected weight difference \( \mathbb{E}[W(i)] - \mathbb{E}[W(i')] \) between a planted biclique node \( i \in \mathcal{A}_0 \) and a non-planted node \( i' \notin \mathcal{A}_0 \).
We compare two scenarios: one where \( i \) is included in the sampled vertex subset \( \mathcal{A} \), and another where \( i' \) is included instead, with the rest of the subset fixed. The weight difference is analyzed by expressing it as a difference of expected terms involving sums over perfect matchings defined by random bijections \( \sigma, \tau \) between \(\mathcal{A}\) and \(\mathcal{B}\).
Conditioning on the planted subsets \((\mathcal{A}_0, \mathcal{B}_0)\), we use Poisson approximations (Lemma~\ref{lem:poisson_intersections}) for the intersection sizes between random subsets and the planted biclique, and a Stein-Chen argument (Lemma~\ref{lem:stein_chen_permutations}) to estimate the overlap of bijections.
The bias arises from differences in exponents involving indicator variables that check whether edges fall within the planted biclique. We enumerate all such configurations and show that the dominant contribution comes from cases where one matching hits the planted structure. This yields a leading term of order \( \Theta(K/n) \), giving the claimed asymptotic bias.
\end{proofsketch}

\begin{corollary}\label{coro:variance-to-weight}
The weights $W(i)$ verify
\begin{align}\label{eq:var_cov_w_i}
&\hbox{Var}(W(i))=\frac{1}{n}\Theta(\dE[W(i)]^2),\\
i\ne j\Rightarrow &\hbox{Cov}(W(i),W(j))=\frac{m^2}{n^2}\Theta(\dE[W(i)]^2).
\end{align}
\end{corollary}
\begin{proof}
    Corollary~\ref{coro:variance-to-weight} follows directly from the definition of joint moment~\eqref{eq:initial_def_phi}, Theorem~\ref{thm:moment} and Proposition~\ref{prop:bias}.
\end{proof}

\subsection{Consequences on Planted Clique Detection}\label{sec:planted-consequences}

Building on the joint moment and bias analysis of vertex weights from the previous subsections, we now discuss their implications for the ability of GBS to detect planted bicliques.

\begin{remark}
Theorem~\ref{thm:moment} indicates that the covariance structures are similar for both biclique and non-biclique nodes, and 
Corollary~\ref{coro:variance-to-weight} establishes that the fluctuation of vertex weight \( W(i) \) is of order \( \Theta\left( \frac{1}{\sqrt{n}} \mathbb{E}[W(i)] \right) \). 
In contrast, Proposition~\ref{prop:bias} shows that the bias in expected weights between biclique and non-biclique nodes scales as \( \Theta\left(\frac{K}{n} \mathbb{E}[W(i)] \right) \). 
Since \( \frac{K}{n} \ll \frac{1}{\sqrt{n}} \) in the regime \( 2\log n \ll K \ll \sqrt{n} \), this implies that the fluctuation scale dominates the signal bias.
Therefore, the frequency with which GBS samples subgraphs containing a given vertex \( i \in V \) does not provide sufficient statistical signal to reliably distinguish biclique nodes from others in this regime.
\end{remark}

In Algorithm~\ref{alg:GBSClique}, the decision function \( f \) can be instantiated with any function based on the vertex weights. 
As a concrete example, we consider a simple function that ranks all vertices in descending order of their weights, selects the top \( c \cdot n \) vertices for a small positive constant $c$, and then applies an existing clique detection algorithm, such as those in~\cite{kuvcera1995expected} or~\cite{alon1998finding}, to the subgraph induced by the selected vertices. 
This approach is motivated by the hope that, due to the weight bias, the planted biclique vertices would appear among those with the largest weights.
If this were the case, identifying the top-weighted vertices could potentially amplify the signal from the planted biclique, thereby increasing the likelihood of success for the downstream detection algorithm.
However, as we demonstrate below, the magnitude of fluctuations overwhelms the bias, rendering this method ineffective for reliably detecting the planted biclique.

\begin{corollary}\label{coro:GBS-usefulness}
    According to Proposition~\ref{prop:bias} and Corollary~\ref{coro:variance-to-weight}, the weights of non-planted/background vertices can be considered as samples from \( f_0 = \mathcal{N}(\mu, \frac{\mu^2}{n}) \), while the weights of planted biclique vertices follow \( f_1 = \mathcal{N}(\mu + \frac{\epsilon \sqrt{n}}{n} \mu, \frac{\mu^2}{n}) \), where $\mu$ is the expected weight of a non-planted vertex and $\epsilon$ is a small positive value. 
    Consider $n-\epsilon\sqrt{n}$ samples from $f_0$ and $\epsilon\sqrt{n}$ samples from $f_1$. 
    Then, the expected number of samples from \( f_1 \) that rank among the top \( c \cdot n \) largest values is approximately
    \begin{align}
        \epsilon\sqrt{n}\left(1- \tilde{\Phi}\left(\tilde{\Phi}^{-1}(1-c)-\epsilon\right)\right), \label{eq:proportion}
    \end{align}
    where $c > 0$ is a small constant and $\tilde{\Phi}$ denotes the CDF of the standard normal distribution.
\end{corollary}

The proof of Corollary~\ref{coro:GBS-usefulness} is presented in Appendix~\ref{sec:proof-coro-GBS-usefulness}. 

\begin{remark}
    Expression~\eqref{eq:proportion} shows that when \( K = \epsilon\sqrt{n} \) is small, the bias is insufficient to significantly elevate the rankings of planted biclique nodes. For example, with \( c = 0.8 \) and \( \epsilon = 0.01 \), only approximately \( 0.8028 \cdot \epsilon\sqrt{n} \) planted biclique nodes rank among the top \( c \cdot n \) vertices -- barely above the baseline proportion \( c \). This implies that GBS-based detection procedure in Algorithm~\ref{alg:GBSClique} provides no significant advantage for planted biclique detection when \( K = o(\sqrt{n}) \). However, when \(\epsilon\) increases, the planted biclique signal becomes more pronounced. For instance, with \( c = 0.8 \) and \( \epsilon = 1 \), roughly \( 0.97 \cdot \epsilon\sqrt{n} \) biclique nodes appear in the top \( c \cdot n \), a proportion substantially exceeding \( c \). This suggests that GBS begins to offer a meaningful advantage for planted clique detection in this higher-signal  \( K = \Theta(\sqrt{n}) \) regime.
\end{remark}

\section{Discussion}\label{sec:discussion}

In this section, we highlight several points that warrant further consideration when interpreting our results. We also present a by-product of our analysis that may be of independent interest.

\paragraph{Planted Clique and Post-Quantum Cryptography.} 
Our finding that GBS-based detection using node frequencies offers no substantial computational advantage for the planted biclique problem in the meaningful regime lends further support to the use of planted clique-based constructions in cryptography~\cite{juels2000hiding, applebaum2010public}. In particular, it enforces their potential as candidates for \emph{post-quantum cryptographic} schemes that remain secure even against quantum adversaries~\cite{bernstein2017post}.
It is important to note that our analysis assumes access to an ideal GBS device. 
In practical implementations, quantum devices are subject to various sources of noise and imperfections, which can further limit the effectiveness of quantum algorithms.

\paragraph{Beyond Node Frequencies in GBS-Based Detection.}
Our analysis has centered on a specific statistic derived from GBS outputs: the frequency with which each node appears in sampled subgraphs, formalized via the notion of {node weights}. This focus enables rigorous theoretical analysis, leading to the conclusion that, under this statistic, the signal-to-noise ratio induced by the planted biclique is too small for reliable detection in the conjectured hard regime.
However, this result does not rule out the potential for more advanced incorporation of GBS samples to succeed in planted biclique detection. Whether alternative GBS-based statistics or algorithms can better exploit its structure remains an open question. While our findings reveal limitations of GBS in this setting, they do not exclude the possibility that other quantum computing paradigms could offer computational advantages for planted clique detection. This remains an open and intriguing direction for future research.

\paragraph{Distribution of Normalized Hafnian in Bipartite Erdös-Rényi Graph.}

As a by-product of our analysis, we characterize the distribution of the Hafnian of a bipartite Erdős–Rényi random matrix $M$ (i.e., a symmetric \((0,1)\)-valued matrix representing the adjacency structure of a random bipartite graph). 
While most prior studies of Hafnians focus on Gaussian random matrices, this result explores a distinct and less commonly analyzed case.
Specifically, we define the normalized Hafnian of $M$ as
\begin{equation}\label{eq:haf-M}
\text{Haf}(M):=\frac{1}{p^n n!}\sum_{\mu\in P_{n}^2}\prod_{\{j, j'\} \in \mu}\xi_{j j'},
\end{equation}
where $P_{n}^2$ denotes the set of all partitions of the set $\{1, ..., n\}$ into subsets of size $2$.
We then show the following distributional convergence:
\begin{theorem}\label{thm:lognormal}
For any fixed $p\in (0,1)$, the normalized Hafnian $\text{Haf}(M)$, as defined in~\eqref{eq:haf-M}, converges weakly as $n\to\infty$ to the Lognormal distribution with parameters  \( \mu = -\frac{1-p}{2p} \) and \( \sigma^2 = \frac{1-p}{p} \). 
\end{theorem}
In other word, Theorem~\ref{thm:lognormal} shows that $\text{Haf}(M)$ converges weakly to $\exp(X)$ where \( X \sim \mathcal{N}(-\frac{1-p}{2p}, \frac{1-p}{p}) \).
This characterization may be of independent interest for further understanding the behavior of GBS on graph problems.
We refer interested readers to Appendix~\ref{sec:proof-thm-lognormal} for the proof of Theorem~\ref{thm:lognormal}.

\section{Conclusion}\label{sec:conclu}

In this work, we investigated whether Gaussian Boson Sampling (GBS) can offer a quantum computational advantage in the classically hard planted biclique detection problem. We focused on a natural GBS-derived statistic, the vertex \emph{weight}, defined as the expected squared number of perfect matchings in all subgraphs of a fixed size that include the given node.

We rigorously analyzed this statistic using probabilistic and combinatorial tools, deriving expressions for the moments of vertex weights. In particular, we first showed that the joint moments of vertex weights converge to those of a multivariate Gaussian, with an explicit covariance structure.
We then quantified the weight bias between planted and non-planted nodes, showing that it scales as $\Theta\left((K/n) \mathbb{E}[W(i)] \right)$.
In the conjectured hard regime where $2\log n \ll K \ll \sqrt{n}$, we proved that this bias is asymptotically negligible compared to the standard deviation $\Theta\left( (1/\sqrt{n}) \mathbb{E}[W(i)] \right) $.
These findings imply that the node occurrence frequency under idealized GBS sampling does not provide a sufficient statistical signal to reliably distinguish planted biclique vertices from others in this regime. While GBS may offer some advantage when $K = \Theta(\sqrt{n})$, its effectiveness appears limited for detecting smaller planted structures.

Overall, our results suggest that although GBS encodes rich combinatorial information via perfect matchings, detecting planted bicliques in the conjectured hard regime remains infeasible under simple weight-based strategies. Future work may explore the performance of more sophisticated GBS-based algorithms or investigate whether other quantum computing paradigms can provide an advantage on the planted (bi)clique problem.

\begin{acknowledgments}
The work of Yu-Zhen Janice Chen is supported by DEVCOM Army Research Laboratory under Cooperative Agreement W911NF-17-2-0196. Laurent Massoulié acknowledges funding support from the PR[AI]RIE-PSAI-ANR-23-IACL-0008 program and the IACL-2023 program. The work of Don Towsley is supported by MURI ARO Grant
W911NF2110325
\end{acknowledgments}

\appendix

\section{Proof of Lemma~\ref{lem:poisson_intersections}}\label{sec:proof-poisson_intersections}

\begin{proof}
Consider vertex subsets $\mathcal{B}_0, \ldots, \mathcal{B}_\ell$ drawn independently uniformly at random from $\binom{V'}{m_j}$ with $m_j = \Theta(\sqrt{n})$, subcollections of vertex subsets $\mathcal{C} \subseteq \{0,\ldots,\ell\}$, and fixed values $x_\cC\in \dN$ denoting the number of elements belong exactly to those vertex subsets in subcollection $\mathcal{C}$.

We denote the number $b_{\{j\}}$ of the nodes $u$ in $V'$ (of size $n$) falling only in vertex subset $\cB_j$ (of size $m_j$), for all $j\in \{0,\ldots,\ell\}$, as 
\begin{align}
    x_{\{j\}} :=m_j - y_j,
\end{align}
where
\begin{align}
    y_j :=\sum_{\cC: j\in \cC, |\cC|\ge 2}x_\cC.
\end{align}
Recall that we denote the total number of nodes shared by at least two vertex subsets as
\begin{align}
    x_{\mathrm{tot}} := \sum_{\mathcal{C} : |\mathcal{C}| \ge 2} x_{\mathcal{C}}.
\end{align}
We then denote the number of nodes that fall into none of the vertex subsets as
\begin{align}
     x_\emptyset=n-z,
\end{align}
where
\begin{align}
    z:=x_{tot}+\sum_{j\in\{0,\ldots,\ell\}} x_{\{j\}}.
\end{align}

Then, the probability that exactly $x_\cC$ nodes of $V'$ appear in the subcollection $\cC$ of vertex subsets, for all $\mathcal{C} \subseteq \{0,\ldots,\ell\}$ (including $\mathcal{C} = \emptyset, \{j\}$) is 
\begin{align}
    &\dP(b_\cC=\displaystyle x_\cC, \forall \cC\subseteq \{0,\ldots,\ell\})\notag\\
    &=\frac{n!}{\prod\limits_{\substack{\cC\subseteq \{0,\ldots, \ell\}}}x_\cC!}\prod_{j=0}^\ell \frac{1}{\binom{n}{m_j}}\\
    &=\displaystyle (n)_z \prod_{j=0}^{\ell}\frac{(m_j)_{y_j}}{(n)_{m_j}} \prod\limits_{\substack{\cC\subseteq \{0,\ldots, \ell\} \\ |\mathcal{C}|\geq 2}}\frac{1}{x_\cC!},
\end{align}
where we introduced the notation
\begin{align}
    (n)_z:=\frac{n!}{(n-z)!},
\end{align}
which can be expressed as
\begin{align}
    (n)_z = n^z \exp\left( \sum_{i=0}^{z-1} \ln\left(1 - \frac{i}{n}\right) \right).
\end{align}

Applying the Taylor expansion for \(\ln(1 - x)\), we obtain the following useful properties:
\begin{align}
    (n)_z &\leq \left(1 + O\left(\frac{z}{n}\right)\right) n^z e^{-z^2 / (2n)}; \\
    z^2 = O(n) \;\Rightarrow\; (n)_z &= \left(1 + O\left(\frac{z}{n}\right)\right) n^z e^{-z^2 / (2n)}. \label{eq:property-2}
\end{align}
In our context, the value of \(z\) satisfies \(z \le \sum_j m_j = O(\sqrt{n})\) by construction. Therefore, we are in the regime where \(z^2 = O(n)\), and~\eqref{eq:property-2} applies. In particular,~\eqref{eq:property-2} allows us to accurately approximate terms like \((n)_{m_j}\) since \(m_j^2 = O(n)\).

In the case where $x_{tot}\le n^{1/4}$, we also have that $y_j^2=O(m_j)$, and we can use the same equivalence to approximate $(m_j)_{y_j}$. In the case where $x_{tot}> n^{1/4}$ we obtain an upper bound instead of an equivalent (we do not explicit this case below).
Thus (for $x_{tot}\le n^{1/4}$):
\begin{align}
&\dP(b_\cC= x_\cC, \forall \cC\subseteq \{0,\ldots,\ell\})\notag\\
&=\displaystyle
\left(1+O\left(\frac{z}{n}+\sum_j \frac{y_j}{m_j} + \sum_j \frac{m_j}{n}\right)\right)\cdot n^{z-\sum_j m_j}\cdot\notag\\
&\qquad\qquad e^{-\frac{z^2}{2n}+\sum_j \frac{m_j^2}{2n} -\sum_j \frac{y_j^2}{2m_j}}\cdot \prod_{j}m_j^{y_j}\prod_{|\cC|\ge 2}\frac{1}{x_\cC!}\\
&\overset{(a)}{=}\displaystyle
\left(1+O\left(\frac{x_{tot}^2}{\sqrt{n}}\right)\right)\cdot n^{z-\sum_j m_j}\cdot\notag\\
&\qquad\qquad e^{-\frac{z^2}{2n}+\sum_j \frac{m_j^2}{2n} -\sum_j \frac{y_j^2}{2m_j}}\cdot \prod_{j}m_j^{y_j}\prod_{|\cC|\ge 2}\frac{1}{x_\cC!}\\
&\overset{(b)}{=}\displaystyle
\left(1+O\left(\frac{x_{tot}^2}{\sqrt{n}}\right)\right)\cdot n^{-\sum_{|\cC|\ge 2}x_\cC(|\cC|-1)}\cdot\notag\\
&\qquad\qquad e^{-\sum_{j<j'}(m_j m_{j'})/n}\prod_{|\cC|\ge 2}\frac{(\prod_{j\in \cC}m_j)^{x_\cC}}{x_\cC !}\\
&\overset{(c)}{=}\displaystyle
\left(1+O\left(\frac{x_{tot}^2}{\sqrt{n}}\right)\right)\cdot\prod_{|\cC|\ge 2}e^{-\mu_\cC}\frac{\mu_\cC ^{x_\cC}}{x_\cC !},
\end{align}
where equality (a) holds because $z \leq x_{tot} \leq n^{1/4}$, $y_j \leq x_{tot}$, $\ell = O(1)$, and $m_j = \Theta(\sqrt{n})$;
equality (b) follows from the definitions of $z, x_{tot}, x_{\{j\}}, y_j$;
finally in equality (c) we artificially introduced factors $e^{-\mu_\cC}$ for $|\cC|\geq2$, which are of order $1+O(\mu_\cC)=1+O(n^{-1/2})$ and are thus accounted for by the factor $(1+O(x_{tot}^2/\sqrt{n}))$.
\end{proof}

\section{Proof of Lemma~\ref{lem:stein_chen_permutations}}\label{sec:proof-stein_chen_permutations}

\begin{proof}
Consider sets of vertices $\cA, \cB, \cA', \cB'$ with $|\cA|=|\cB|$, $|\cA'|=|\cB'|$,  
\begin{align}
    |\cA|,|\cB|,|\cA'|,|\cB'|,|\cA\cap \cA'|, |\cB\cap \cB'|=m-o(m),
\end{align}
and independent bijections $\sigma$, $\tau$ sampled uniformly at random from respectively $\hbox{Bij}(\cA, \cB)$ and $\hbox{Bij}(\cA', \cB')$.
In the following, we analyze the probability that the two bijections map to different vertices for all overlapping vertices using the Stein-Chen method~\cite[\S 3.3]{draief2010epidemics}. 

To that end, we define $Z_i :=\II_{\sigma(i)=\tau(i)}$, $i\in \cA\cap \cA'$. 
In addition, for any fixed $i \in \cA \cap \cA'$, we define modified bijections $\sigma'$, $\tau'$ as follows. 
\begin{itemize}
    \item We sample node $v$ uniformly at random from $\cB\cap \cB'$.
    \item Then, we let $\sigma'(i)=v$, and for $u:=\sigma^{-1}(v)$ we let $\sigma'(u)=\sigma(i)$. 
    \item Similarly, we let $\tau'(i)=v$, and for $u':=\tau^{-1}(v)$ we let $\tau'({u'})=\tau(i)$. 
\end{itemize}
Obviously, the pair $(\sigma',\tau')$ has the same distribution as $(\sigma,\tau)$ conditioned on $\sigma(i)=\tau(i)$. 

We then let for all $j\in \cA\cap \cA'\setminus \{i\}$:
\begin{align}
    Z_{j|i}:=\II_{\sigma'(j)=\tau'(j)}.
\end{align}
By construction, the distribution of $\{Z_{j|i}\}_{j\in \cA\cap \cA'\setminus\{i\}}$ coincides with the distribution of $\{Z_{j}\}_{j\in A\cap A'\setminus\{i\}}$ conditionally on $Z_i=1$. 
Thus, Stein-Chen's lemma applies to give the following.
Denote 
\begin{align}
    \pi_i &:= \dE [Z_i]= \sum_{v \in\cB \cap \cB'} \dP(\sigma(i) = v) \dP(\tau(i) = v)
    =\frac{|\cB \cap \cB'|}{|\cB| \cdot |\cB'|},\\
    Z&:=\sum_{i\in \cA\cap \cA'} Z_i,\\
    \lambda &:=\dE [Z]=\sum_{i\in \cA\cap \cA'} \pi_i=\frac{|\cA\cap \cA'||\cB\cap \cB'|}{|\cB|\cdot |\cB'|}=1+o(1).
\end{align}
We then have
\begin{align}
    &d_{\hbox{Var}}(Z,\hbox{Poi}(\lambda))\notag\\
    &\le 2\min(1,\lambda^{-1})\sum_{i\in \cA\cap \cA'}\pi_i\left(\pi_i+\sum_{j\ne i}\dE\left[|Z_j-Z_{j|i}|\right]\right).\label{eq:Stein_Chen_distinct_sets}
\end{align}

Note that, for $j\in \cA \cap \cA' \setminus \{i\} $, $\sigma(j)=\sigma'(j)$ unless $\sigma(j)=v$, and similarly $\tau(j)=\tau'(j)$ unless $\tau(j)=v$; hence, $Z_j \neq Z_{j|i}$ only if $j = \sigma^{-1}(v)$ or $j = \tau^{-1}(v)$.
Thus,
\begin{align}
    &\sum_{j\ne i}\dE\left[|Z_j-Z_{j|i}|\right]\notag\\
    &\le\dE \left[|Z_{\sigma^{-1}(v)}-Z_{\sigma^{-1}(v)|i}|\right]+\dE\left[|Z_{\tau^{-1}(v)}-Z_{\tau^{-1}(v)|i}|\right] \\
    &\le \dE\left[Z_{\sigma^{-1}(v)}+Z_{\sigma^{-1}(v)|i}+Z_{\tau^{-1}(v)}+Z_{\tau^{-1}(v)|i}\right]\\
    &\le 2\frac{|\cB\cap \cB'|}{|\cB|\cdot |\cB'|}+2\frac{|\cB\cap \cB'|-1}{(|\cB|-1)\cdot (|\cB'|-1)}\\
    &=O\left(\frac{1}{m}\right). \label{eq:Zj-Zji}
\end{align}
Plugging~\eqref{eq:Zj-Zji} in~\eqref{eq:Stein_Chen_distinct_sets} yields
\begin{align}
    &d_{\hbox{Var}}(Z,\hbox{Poi}(\lambda))\notag\\
    &\le 2 \min(\lambda, 1)\left[\frac{\lambda}{|\cA\cap \cA'|}+O\left(\frac{1}{m}\right)\right]\\
    &=O\left(\frac{1}{m}\right).
\end{align}

This shows that the random variable $Z$ behaves approximately like a Poisson random variable with mean $1+o(1)$. 

In particular, 
\begin{align}
    \dP(Z=0)-e^{-\lambda}=O\left(\frac{1}{m}\right),
\end{align}
so that, since $\lambda=1+o(1)$,
\begin{align}
    \dP(Z=0)=e^{-1}+o(1),
\end{align}
which proves \eqref{eq:lemma_2_eq1}.
Write then
\begin{align}\label{pzk0}
    \mathbb{E}\left(p^{-Z}\right) &=\sum_{k=0}^m p^{-k}\mathbb{P}(Z=k) .
\end{align}
Now,
\begin{align}\label{eq:pzk}
 \mathbb{P}(Z=k)    &=\sum_{\tilde A\in \binom{\cA\cap \cA'}{k}}\sum_{\tilde B\in \binom{\cB\cap\cB'}{k}}\frac{k!}{(m)_k (m)_k}\dP(\tilde Z=0),
    \end{align}
where we introduced the sets $\tilde A$, $\tilde B$ such that $\sigma$ and $\tau$ agree on $\tilde A$, and are such that $\sigma(\tilde A)=\tilde B$. Letting $\tilde \sigma$ and $\tilde\tau$ be two independent uniformly random bijections from  $\text{Bij}(\cA\setminus \tilde \cA, \cB \setminus \tilde \cB)$ and $\text{Bij}(\cA'\setminus \tilde A, \cB'\setminus \tilde B)$ respectively, in the above expression we have  $\tilde Z := \sum_{u\in\cA\setminus \tilde A}\II_{\tilde \sigma(u)=\tilde \tau(u)}$.

Now thanks to \eqref{eq:lemma_2_eq1}, we know that for any fixed $k$, $\dP(\tilde Z=0)=e^{-1}(1+o(1))$. Thus, 
\begin{align}\label{eq:pzk2}
 \mathbb{P}(Z=k)    &=\frac{|\cA\cap \cA'|_k |\cB\cap \cB'|_k}{(m)_k (m)_k}\frac{e^{-1}}{k!}(1+o(1))\\
 &=(1+o(1)) \frac{e^{-1}}{k!},
    \end{align}
where we used that $|\cA\cap \cA'|,|\cB\cap \cB'|=m-o(m)$. In addition, the previous expression \eqref{eq:pzk} implies that $\dP(Z=k)\le \frac{1}{k!}$. This together with \eqref{eq:pzk2} and \eqref{pzk0} implies, by Lebesgue's dominated convergence, that 
\begin{align}
    \mathbb{E}\left(p^{-Z}\right) &=(1+o(1))\sum_{k=0}^{\infty} p^{-k}\frac{e^{-1}}{k!},
\end{align}
thus establishing \eqref{eq:lemma_2_eq1}.
\end{proof}

\section{Proof of Theorem~\ref{thm:moment}}\label{sec:proof-thm-moment}

\begin{proof}
We prove Theorem~\ref{thm:moment} in the following four steps.

\subsection{\textbf{Step 1: Reformulate moment expression $\phi(i(1),\ldots,i(\ell))$.}}
For each \( j \in [\ell] \), let \( \mathcal{A}_j \subseteq V \) be a random subset of size \( m \), sampled uniformly among all such subsets that contain the vertex \( i(j) \). Let \( \mathcal{B}_j \subseteq V' \) be a uniformly random subset of size \( m \). Let \( \sigma^j \) and \( \tau^j \) be two independent bijections sampled uniformly from \( \mathrm{Bij}(\mathcal{A}_j, \mathcal{B}_j) \).

We define the following notations to assist the analysis of the joint moment defined by~\eqref{eq:initial_def_phi},~\eqref{eq:W}, and~\eqref{eq:W_expectation}.
For each \( j \in [\ell] \), we denote by $E_{j,\sigma}$ (resp., $E_{j,\tau}$) the set of matching edges from $\cA_j$ to $\cB_j$ via $\sigma^j$ (resp., $\tau^j$), i.e., 
\begin{align}
    E_{j,\sigma} &:= \{e=(u,\sigma^j(u)),u\in\cA_j\},\\
    E_{j,\tau} &:= \{e=(u,\tau^j(u)),u\in\cA_j\}.
\end{align}
We further denote by $E_j$ all edges from both matchings between $\cA_j$ and $\cB_j$, i.e.
\begin{align}
    E_j:=E_{j,\sigma}\cup E_{j,\tau},
\end{align}
denote by $E'_j$ the set of edges in $E_j$ excluding those fully within the planted biclique $\cA_0 \times \cB_0$, i.e.,
\begin{align}
    E'_j :=E_j \setminus \{\cA_0\times \cB_0\},
\end{align}
denote by $F_j$ the set of edges in $E'_j$ that are shared by some other $E'_k$, for $k \in [\ell]$, $k \neq j$, i.e.,
\begin{align}
    F_j:= E'_j \bigcap \left\{\cup_{k\in [\ell]\setminus\{j\}}E'_k\right\},
\end{align}
and denote by $G_j$ the set of edges in $E'_j$ that are unique to $j$ (not found in any other $E'_k$), i.e., 
\begin{align}
    G_j:=E'_j\setminus F_j.
\end{align}
Finally, for each edge $e\in E_j$, we denote by $k_{e,j} \in \{0, 1, 2\}$ the number of times edge $e$ appears in $E_j$ (e.g., never, once in either matching, or twice in both matchings), i.e., 
\begin{align}
    k_{e,j}:=\II_{e\in E_{j,\sigma}}+\II_{e\in E_{j,\tau}}.
\end{align}

Using these new notations, we rewrite~\eqref{eq:W} to
\begin{align}
    W(i(j)) &= \mathbb{E}_{(\cA, \cB, \sigma, \tau)_j} \left[p^{-2m}\prod_{e \in E_j}\xi_{e}\right]\\ &= \mathbb{E}_{(\cA, \cB, \sigma, \tau)_j} \left[p^{-2m}\prod_{e \in E'_j}\xi_{e}\right],
\end{align}
rewrite~\eqref{eq:W_expectation} to 
\begin{align}
    \mathbb{E}[W(i(j))] 
    &= \mathbb{E}\left[p^{-2m}\prod_{e \in E'_j, }p\right],
\end{align}
and rewrite~\eqref{eq:initial_def_phi} to
\begin{align}
    \phi(i(1),\ldots,i(\ell))=n^{\ell/2}\dE \left[\prod_{j\in [\ell]} p^{-2m}\left((\prod_{e\in E'_j} \xi_e) -p^{|E'_j|}\right)\right].
\end{align}

We can evaluate this expectation by first conditioning over the choices $\cA_j$, $\cB_j$, $\sigma^j$, $\tau^j$, and integrating over the $\xi_e$. Remark that this conditional expectation is equal to zero unless for all $j\in [\ell]$, one has $F_j\ne \emptyset$. 
We thus have:
\begin{align}\label{eq:conditional_expectation_weight_moments}
&\phi(i(1),\ldots,i(\ell))\notag\\
&=n^{\ell/2}\dE\left[\II_{F_j\ne \emptyset \forall j}
\prod_{j\in[\ell]}p^{-2m +|G_j|}\left( \prod_{e\in F_j}\xi_e -p^{|F_j|}\right)\right].
\end{align}
The exponent $-2m+|G_j|$ also reads
\begin{align}
    &-2m+|G_j|\notag\\
    &=-2m+|E_j|-|E_j\cap \cA_0\times \cB_0|-|F_j|\\
    &=-2m+2m-\sum_{u\in \cA_j}\II_{\sigma^j(u)=\tau^j(u)}-|E_j\cap \cA_0\times \cB_0|-|F_j|.
\end{align}
Thus, we are left with the evaluation of 
\begin{align}
&\phi(i(1),\ldots,i(\ell)) = n^{\ell/2}\dE\left[\II_{F_j\ne \emptyset,  \forall j} \cdot \phi'\right], \label{eq:conditional_expectation_weight_moments_bis}
\end{align}
where
\begin{align}
    &\phi' := \prod_{j\in[\ell]}p^{-(\sum_{u\in \cA_j}\II_{\sigma^j(u)=\tau^j(u)}+|E_j\cap \cA_0\times \cB_0|+|F_j|)}\notag\\
    &\qquad\qquad\qquad\qquad\qquad \times\left( \prod_{e\in F_j}\xi_e -p^{|F_j|}\right) \label{eq:phi'}
\end{align}

To evaluate this expectation, we need more notation. 
We let 
\begin{align}
    \cD &:=\{i(1),\ldots,i(\ell)\},\\
    \cD_j &:=\cD\cap \cA_j, \quad \forall j\in[\ell].
\end{align}
Thus, by construction $i(j)\in \cD_j$. For any subset $\cD'\subset \cD$ that contains $i(j)$, the probability that $\cD_j=\cD'$ equals
\begin{align}\label{eq:D'}
\frac{\binom{n-|\cD|}{m-|\cD'|}}{\binom{n-1}{m-1}}=\frac{(n-|\cD|)_{m-|\cD'|}(m-1)_{|\cD'|-1}}{(n-1)_{m-1}},
\end{align}
where we introduced the notation
\begin{align}
    (n)_z:=\frac{n!}{(n-z)!},
\end{align}  
and conditioned on that event, the set 
\begin{align}
    \cE_j:=\cA_j\setminus \cD
\end{align}
is chosen uniformly at random from $\binom{V\setminus\cD}{m-|\cD_j|}$.

For each subset $\cC\subset [\ell]$, we let 
\begin{align}
    A_{ \cC}&:=\cap_{j\in \cC}\cA_j \setminus \{\cup_{j\notin \cC}\cA_j\}, &a_\cC&:=|A_{ \cC}|,\label{eq:ac}\\
    B_{ \cC}&:= \cap_{j\in \cC}\cB_j \setminus \{\cup_{j\notin \cC}\cB_j\}, &b_\cC&:=|B_{ \cC}|,\label{eq:bc}\\
    D_{ \cC}&:= \cap_{j\in \cC}\cD_j \setminus \{\cup_{j\notin \cC}\cD_j\}, &d_\cC&:=|D_{ \cC}|,\label{eq:dc}\\
    E_{ \cC}&:=\cap_{j\in \cC}\cE_j \setminus \{\cup_{j\notin \cC}\cE_j\}, &e_\cC&:=|E_{ \cC}|.\label{eq:ec}
\end{align}
By construction, $\cA_{\cC}=\cD_{\cC}\cup\cE_{\cC}$ and $a_\cC=d_\cC+e_\cC$. 

Let, for each $j\in [\ell]$, $f_{j,\sigma}$ (respectively, $f_{j,\tau}$) denote the number of edges of $E_{j,\sigma}$ (respectively, $E_{j,\tau}$) that are shared with or lie inside the regions defined by some other cross-product sets, $\cA_k \times \cB_k, k \neq j$, i.e., 
\begin{align}
    f_{j,\sigma} & := E_{j, \sigma} \bigcap \left\{\cup_{k \in \{0, ..., \ell\} \setminus \{j\}}\cA_k \times \cB_k \right\}, \\
    f_{j,\tau} & := E_{j, \tau} \bigcap \left\{\cup_{k \in \{0, ..., \ell\} \setminus \{j\}}\cA_k \times \cB_k \right\}.
\end{align}
Let, for each $j\in [\ell]$, $a_j$ (resp., $b_j$) denote the number of nodes of $\cA_j$ (resp., $\cB_j$) that are shared with some other vertex subsets, $\cA_k$, $k \neq j$, i.e., 
\begin{align}
    a_j &:=\left|\cA_j \cap \cup_{k \in \{0, ..., \ell\} \setminus \{j\}}\cA_{k}\right|, \\
    b_j &:=\left|\cB_j \cap \cup_{k \in \{0, ..., \ell\} \setminus \{j\}}\cB_{k}\right|.
\end{align}

Condition of the size $a_j$, $b_j$, the number of elements in $f_{j,\sigma}$ can be seen as a hypergeometric random variable with parameters $(m, a_j, b_j)$ by viewing the number of elements as the number of balls falling into $a_j$ marked urns when throwing $b_j$ balls into distinct urns from a total collection of $m$ such urns, which is stochastically less than a binomial random variable with parameters $(b_j, a_j/(m-b_j))$.
Note that $|F_j|\le f_{j,\sigma}+f_{j,\tau}$, and that $f_{j,\sigma}$, $f_{j,\tau}$ are identically distributed and independent conditioned on $a_j,b_j$.
Thus, conditioned on $a_j,b_j$, 
\begin{align}
    |F_j|\le f_j:=f_{j,\sigma}+f_{j,\tau}\le \hbox{Bin}(2b_j, a_j/(m-b_j)).
\end{align}

Using this, let us show that the leading contribution in \eqref{eq:conditional_expectation_weight_moments_bis} is obtained when restricting oneself to the event on which $|F_j|=1, \forall j\in[\ell]$. 
Indeed, the left-hand side in \eqref{eq:conditional_expectation_weight_moments_bis} equals $\phi_1+\phi_2$ where 
\begin{align}
\phi_1&:=n^{\ell/2}\dE\left[\II_{f_j=1\&|F_j|=1, \forall j}
 \cdot \phi'\right], \label{eq:phi_1et2}\\ 
\phi_2&:=n^{\ell/2}\dE\left[\II_{F_j\ne \emptyset, \forall j,\sum_j f_j>\ell}\cdot \phi'\right],
\end{align}
and $\phi'$ is defined in~\eqref{eq:phi'}.

\subsection{\textbf{Step 2: Show that contribution of multiple-edge overlaps $\phi_2$ is negligible.}}
Noting that the absolute value of $\prod_{e\in F_j}\xi_e -p^{|F_j|}$ is less than one, we have
\begin{align}
    |\phi_2| &\le n^{\ell/2}\dE\Bigg[\II_{F_j\ne \emptyset, \forall j,\sum_j f_j>\ell}\times\notag\\
    &\qquad\prod_{j\in[\ell]}p^{-(\sum_{u\in \cA_j}\II_{\sigma^j(u)=\tau^j(u)}+|E_j\cap \cA_0\times \cB_0|+|F_j|)}\Bigg]. \label{eq:phi2}
\end{align}

\subsubsection{\textbf{Step 2.a: Introduce conditional matching construction procedure.}}
In the following, we describe a conditional matching construction procedure, which later on will be used for upper bounding $\phi_2$.

Given two vertex sets $\cA, \cB$ of size $m$ and two small subsets of them $\cA' \subset \cA$, $\cB' \subset \cB$, we uniformly randomly pick a bijection $\sigma \in \hbox{Bij}(\cA,\cB)$ and let $\cF$ denote the collection of edges $(u,\sigma(u))$ of the perfect matching  between $\cA$ and $\cB$ induced by $\sigma$ that belong to $\cA'\times \cB'$, i.e.,  
\begin{align}
    \cF:=\{e=(u,\sigma(u)): u\in \cA', \sigma(u)\in \cB'\}.
\end{align}
We further uniformly randomly pick another bijection $\sigma' \in \hbox{Bij}(\cA,\cB)$ and let initially $\sigma'' = \sigma'$.

We now describe the two-phase construction of the remainder of the matching, conditioned on $\cF$.

\textit{Phase 1: } In this phase, we modify $\sigma''$ such that the intersection between the matching induced by $\sigma''$ and $\cA'\times \cB'$ contains $\cF$. 
Denote by $\{(u_1,v_1)\ldots,(u_{|\cF|},v_{\cF|})\}$ the edges in $\cF$. 
Sequentially for each $(u_s,v_s)\in \cF$, do the following.
{
\begin{itemize}
    \item If $\sigma''({u_s})=v_s$, then do nothing.
    \item If $\sigma''({u_s})=v' \ne v_s$,  then let $u'=(\sigma'')^{-1}(v_s)$ and let $\sigma''({u_s})=v_s$, $\sigma''({u'})=v'$.
\end{itemize}
}
It is easily shown that after each step $s$, permutation $\sigma''$ is uniform over $\hbox{Bij}(\cA,\cB)$ conditional on $\sigma''({u_t})=v_t$, $t\in [s]$.
Furthermore, at the end of this phase, at most $2|\cF|$ edges of $\sigma''$ have been modified.

\textit{Phase 2: } In this phase, we further modify $\sigma''$ such that, for all $u \in \cA', (u, \sigma(u)) \notin \cF$, we make $\sigma''(u) \notin \cB'$. Toward this, we order the vertices $u$ in $\cA'$ that are not the origin of an edge in $\cF$ and, without loss of generality, assume that their indices are the integers in $[L]$ for some $L$.
Sequentially for each $u\in  [L]$, do the following.
\begin{itemize}
    \item If $\sigma''(u)\notin \cB'$, then do nothing.
    \item If $\sigma''(u)=v \in \cB'$, then select uniformly at random some $v'\in \cB\setminus \cB'$ such that $u':=(\sigma'')^{-1}(v')\notin [u-1]$ and modify $\sigma''$ by letting $\sigma''(u)=v'$, $\sigma''({u'})=v$.
\end{itemize}
It is easily shown that at the end of stage $u$ of this phase, permutation $\sigma''$ is uniformly distributed over $\hbox{Bij}(\cA,\cB)$ conditionally on $\sigma''({u_s})=v_s$ for all $s\in [\cF]$, and $\sigma''(i)\notin \cB'$ for all $i\in [u]$.
Moreover, at the end of this phase, at most $2L=2(|\cA'|-|\cF|)$ edges of $\sigma''$ have been modified. 

Finally, this conditional matching construction procedure outputs bijection $\sigma''$ from the distribution of $\sigma$ conditionally on the intersection of the induced matching with $\cA'\times \cB'$ by modifying at most $2|\cA'|$ edges of $\sigma'$, where $\sigma'$ was independent of $\sigma$. 

\subsubsection{\textbf{Step 2.b: Bound $\phi_2$ using Step 2.a, binomial-type argument, and Poisson tail bound.}}
In the following, we use this construction procedure to upper-bound $\phi_2$. 
Specifically, for each $j\in[\ell]$, we consider $\sigma'^j$, $\tau'^j$ sampled uniformly from $\hbox{Bij}(\cA_j,\cB_j)$ independently of everything else, and $\sigma''^j$, $\tau''^j$ their modification according to the above construction (with $\cA' = \cA_j \cap \cup_{k \in \{0, ..., \ell\} \setminus \{j\}}\cA_{k}$ and $\cB' = \cB_j \cap \cup_{k \in \{0, ..., \ell\} \setminus \{j\}}\cB_{k}$).
Then, we have
\begin{align}
    \sum_{u\in \cA_j}\II_{\sigma''^j(u)=\tau''^j(u)}\le 4 a_j +\sum_{u\in \cA_j}\II_{\sigma'^j(u)=\tau'^j(u)}.
\end{align}
Note also, by definitions, 
    \(|E_j\cap \cA_0\times \cB_0|\le f_j. \)
These, together with~\eqref{eq:phi2}, give:
\begin{align}
    |\phi_2|&\le n^{\ell/2}\dE\Bigg[\II_{f_j \ge 1 \forall j\&\sum_{j\in[\ell]}f_j\ge \ell+1} \times\notag\\
    &\qquad \prod_{j\in [\ell]}p^{-(\sum_{u\in \cA_j}\II_{\sigma'^j(u)=\tau'^j(u)}+4  a_j+ 2 f_j)}\Bigg].
\end{align}

Lemma \ref{lem:stein_chen_permutations}, Equation \eqref{eq:lemma_2_eq2} entails that the expectation of the factors $p^{-\sum_{u\in \cA_j}\II_{\sigma'^j(u)=\tau'_j(u)}}$ conditionally on the sets \( \mathcal{A}_j \), \( \mathcal{B}_j \) equals $O(1)$.

Moreover, we can further upper-bound $f_j$ by $2 a_j$.
We finally get
\begin{align}
    |\phi_2|\le O(n^{\ell/2})\dE\left[\II_{f_j \ge 1, \forall j, \sum_{j\in[\ell]}f_j\ge \ell+1} \cdot p^{-8\sum_{j\in[\ell]}a_j}\right].
\end{align}

Write next
\begin{align}
    \II_{(f_j \ge 1, \forall j)\&(\sum_{j\in [\ell]} f_j\ge \ell+1)}\le \sum_{j}\II_{(f_k\ge 1,\forall k\neq j)\& (f_j \ge 2)}.
\end{align}
The probability that a binomial random variable with parameters $(a,q)$ is at least $1$ is upper-bounded by its expectation, $aq$ (Markov's inequality), and the probability that it is at least $2$ is upper-bounded by $ a^2q^2$.
Thus, the expectation of the summation in the previous display,  conditionally on the $\cA_j,\cB_j$, is upper-bounded by
\begin{align}
    \sum_j \frac{4a_j^2 b_j^2}{(m-b_j)^2}\prod_{k\ne j}\frac{2 a_k b_k}{m-b_k}.
\end{align}

Finally, we let $b_\cC$ be the size of $B_\cC$ (see Lemma \ref{lem:poisson_intersections}) for $\cC\subset \{0,\ldots,\ell\}$, $|\cC\ge 2|$, and distinguish according to whether $b_{tot}\le n^{1/4}$ or not ($b_{tot} := \sum_{\cC: |\cC| \geq 2} b_{\cC}$, which is $\ll m$). When $b_{tot}\le n^{1/4}$, then $m-b_j\ge m/2$ for all $j\in [\ell]$. Thus:
\begin{align}
    |\phi_2| &\le O(n^{\ell/2})\dE\left[\left(\sum_j \frac{4a_j^2 b_j^2}{(m/2)^2}\prod_{k\ne j}\frac{2 a_k b_k}{m/2} \right) p^{-8\sum_{j\in[\ell]}a_j}   \right]\notag\\  &\qquad+O(n^{\ell/2})\dE\left[\II_{X_{tot}> n^{1/4}}\cdot p^{-8\sum_{j\in[\ell]}a_j}\right]\\
    &\overset{(a)}{\le} O(n^{-1/2})\dE\left[ \left(\sum_j 4a_j^2 b_j^2\prod_{k\ne j}2 a_k b_k \right) p^{-8\sum_{j\in[\ell]}a_j}   \right]\notag\\
    &\qquad+O(n^{\ell/2})\dP(X_{tot} >n^{1/4})\dE\left[p^{-8\sum_{j\in[\ell]}a_j}   \right]
\end{align}
where in inequality (a) we use that $m=\Theta(n^{1/2})$ to bound the first term.

Note, in view of Lemma \ref{lem:poisson_intersections} that
\begin{align}
    &\dE\left[ \left(\sum_j 4a_j^2 b_j^2\prod_{k\ne j}2 a_k b_k \right) p^{-8\sum_{j\in[\ell]}a_j}   \right]=O(1), \\ &\dE\left[p^{-8\sum_{j\in[\ell]}a_j}   \right]=O(1),
\end{align}
so that
\begin{align}
    |\phi_2|\le O(n^{-1/2})+ O(n^{\ell/2})\dP(X_{tot}>n^{1/4}).
\end{align}
Using again Lemma \ref{lem:poisson_intersections}, we get an upper bound in the above if we replace $X_{tot}$ by a Poisson random variable with parameter
\begin{align}
    \lambda:=\sum_{\cC} \mu_{\cC}=O(1),
\end{align}
where the summation is over $\cC\subset \{0,\ldots, \ell\}$ of size at least 2. Now we may use that for large enough $n$, 
\begin{align}
    \dP(\hbox{Poi}(\lambda)\ge n^{1/4})\le e^{-n^{1/4}}\ll n^{-\ell/2 -1/2}
\end{align}
to conclude:
\begin{align}
|\phi_2|= O(n^{-1/2}).
\end{align}

\subsection{\textbf{Step 3: Show that leading contribution comes from single-edge overlaps $\phi_1$.}}
It remains to evaluate $\phi_1$ as defined in \eqref{eq:phi_1et2}. 

Let us define $\cM$ to be the event that each $F_j$ is reduced to a single edge, and that this single edge belongs to $F_{j'}$ for only one $j'$ in $[\ell]\setminus\{j\}$, i.e.,
\begin{align}
    \cM := \{\forall j \in [\ell], |F_j| = 1, \exists !j' \in [\ell] \setminus\{j\}, F_j = F_{j'} \}.
\end{align}
On event $\cM$, we denote by $M$ the perfect matching between the elements of $[\ell]$ induced by the relation 
$F_j\cap F_{j'}\ne\emptyset$. For $\{j,j'\}$ a pair in this matching $M$.

Write then $\phi_1$  as $\psi_1+\psi_2$, where
\begin{align}
    \psi_1&\displaystyle :=n^{\ell/2}\dE\left[\II_{f_j=1 \forall j}\II_\cM \cdot \phi'\right],\\
    \psi_2&\displaystyle :=n^{\ell/2}\dE\left[\II_{f_j=1\&|F_j|=1 \forall j}\II_{\overline{\cM}}\cdot \phi'\right]. \label{eq:psi_1_et_2}
\end{align}

\subsubsection{\textbf{Step 3.a: Show that contribution of overlaps without perfect matching $\psi_2$ is negligible. }}
Let us first deal with $\psi_2$. Using similar arguments as we used to bound $|\phi_2|$, we have
\begin{align}
    |\psi_2|\le O(n^{\ell/2})\dE\left[\II_{f_j=1\&|F_j|=1\forall j}\II_{\overline{\cM}} \cdot p^{-8\sum_j a_j}\right].
\end{align}
We may again distinguish according to whether $X_{tot}\le n^{1/4}$ or not, the latter term leading here also to an $O(n^{-1/2})$ contribution. Now the only way one can have $|F_j|=1$ for all $j$ and not be in $\cM$ is when some edge belongs to at least three distinct sets $F_j$, $F_{j'}$ and $F_{j''}$. This can only happen if for some set $\cC$ of size $|\cC|\ge 3$, one has $|\cB_\cC|\ge 1$. 
Thus,
\begin{align}
    |\psi_2|&\le O(n^{-1/2})+O(n^{\ell/2})\times\notag\\
    &\qquad\dE\left[\II_{f_j\ge 1\forall j}\II_{X_{tot}\le n^{1/4}}\II_{\sum_{|\cC|\ge 3}b_{\cC}\ge 1} \cdot p^{-8\sum_j a_j}\right]\\
    &\le O(n^{-1/2})+O(n^{\ell/2})\times\notag\\
    &\qquad\dE\left[\left(\prod_j\frac{2a_j b_j}{m/2}\right)\II_{\sum_{|\cC|\ge 3}b_{\cC}\ge 1} \cdot p^{-8\sum_j a_j}\right]\\
    &\le O(n^{-1/2})+O(1)\dE\left[\left(\prod_j  b_j\right)\II_{\sum_{|\cC|\ge 3}b_{\cC}\ge 1}\right],
\end{align}
where we used $m=\Theta(n^{1/2})$, independence between $b_\cC$ and $a_j$ and Lemma \ref{lem:poisson_intersections} to bound by $O(1)$ the expectation of the term $(\prod_j a_j) p^{-8\sum_j a_j}$. 

Write then (relying again on Lemma \ref{lem:poisson_intersections})
\begin{align}
    &\dE\left[\left(\prod_j b_j\right)\II_{\sum_{|\cC|\ge 3}b_{\cC}\ge 1}\right] \notag\\
    &\le O(1)\sum_{|\cC|\ge 3} \dE[ b_{\cC}\II_{b_\cC\ge 1}]\\
    &\le O(1)\sum_{|\cC|\ge 3} \mu_\cC.
\end{align}
By definition of $\mu_\cC$, the right-hand side of this last expression is $O(n^{-1/2})$. We thus conclude that 
\begin{align}
|\psi_2|=O(n^{-1/2}).
\end{align}

\subsubsection{\textbf{Step 3.b: Identify that leading contribution comes from configuration $\psi_1$ where shared edges form a perfect matching.}}
It remains to evaluate $\psi_1$.
Note that necessarily, $\psi_1=0$ for odd $\ell$ since no perfect matching exists on an odd-sized set.
Write
\begin{align}
    \psi_1&\displaystyle =n^{\ell/2}\dE\Bigg[\II_{f_j=1\forall j}\II_\cM \prod_{e=\{j,j'\}\in M}(\xi_{e}-p)^2\times\notag\\
    &\qquad \prod_{j\in[\ell]}\left(\frac{1}{p}\right)^{\sum_{u\in \cA_j}\II_{\sigma^j(u)=\tau^j(u)}+|E_j\cap \cA_0\times \cB_0|+1}\Bigg]\\
    &=\displaystyle n^{\ell/2} p^{-\ell}[p(1-p)]^{\ell/2}\dE\Bigg[\II_{f_j=1\forall j}\II_\cM \times\notag\\ 
    &\qquad \prod_{j\in[\ell]}\left(\frac{1}{p}\right)^{\sum_{u\in \cA_j}\II_{\sigma^j(u)=\tau^j(u)}+|E_j\cap \cA_0\times \cB_0|}\Bigg]\\
    &=\displaystyle n^{\ell/2} p^{-\ell}[p(1-p)]^{\ell/2}\sum_{\mu}\dE\Bigg[\II_{f_j=1\forall j}\II_\cM\II_{M=\mu}\times\notag\\ 
    &\qquad\prod_{j\in[\ell]}\left(\frac{1}{p}\right)^{\sum_{u\in \cA_j}\II_{\sigma^j(u)=\tau^j(u)}+|E_j\cap \cA_0\times \cB_0|}\Bigg].
\end{align}

In the above, the summation is over all perfect matchings $\mu$ of $[\ell]$, i.e., over the set of all partitions of the set $\{1, ..., \ell\}$ into subsets of size $2$.

Note that on the event $\cM$, for each $j\in[\ell]$ there is only one edge in $F_j$, which by definition of $F_j$  cannot belong to $\cA_0\times\cB_0$; on the event $\{f_j=1\}\cap \cM$, therefore there is no edge  in $E_j \cap \cA_0\times \cB_0$. We may thus drop the exponent $|E_j \cap \cA_0\times \cB_0|$ in the last display to obtain
\begin{align}
    \psi_1&=n^{\ell/2} p^{-\ell}[p(1-p)]^{\ell/2}\sum_{\mu}\dE\Bigg[\II_{f_j=1\forall j}\II_\cM\II_{M=\mu}\times \notag\\
    &\qquad\qquad\qquad\prod_{j\in[\ell]}\left(\frac{1}{p}\right)^{\sum_{u\in \cA_j}\II_{\sigma^j(u)=\tau^j(u)}}\Bigg].\label{eq:psi_1}
\end{align}
Using the same arguments as in the control of $\psi_2$, we may include into the previous expectation defining $\psi_1$ the indicator $\II_\cN$ of the event $\cN$ defined as:
\begin{align}
    \cN=\left\{\sum_{|\cC|\ge 3} |\cB_\cC|=0\; \& \sum_{|\cC|\ge 3} |\cE_\cC|=0\right\}.
\end{align}

To determine the limit of $\psi_1$ as $n\to\infty$, we evaluate upper and lower bounds of the terms in the summation over matchings $\mu$.

\paragraph{\textbf{Step 3.b.a: Evaluate Upper Bound.}}
Write for each perfect matching $\mu$ of $[\ell]$:
\begin{align}\label{eq:psi_1_upper}
\dE\left[\II_{f_j=1\forall j}\II_\cN\II_\cM\II_{M=\mu} \prod_{j\in[\ell]}\left(\frac{1}{p}\right)^{\sum_{u\in \cA_j}\II_{\sigma^j(u)=\tau^j(u)}}\right]\le H_\mu,
\end{align}
where
\begin{align}
    H_\mu &=\displaystyle \dE\Bigg[\II_{\cN} \sum_{\substack{(u_e,v_e)\in \cA_j\cap\cA_{j'}\times \cB_j\cap \cB_{j'}, e=\{j,j'\}\in \mu}}\notag\\
    &\prod_{e=\{j,j'\}\in\mu}\Bigg(\II_{\sigma^j({u_e})=\sigma^{j'}({u_e})=v_e}+\II_{\sigma^j({u_e})=\tau^{j'}({u_e})=v_e}\notag\\
    &\quad+\II_{\tau^j({u_e})=\sigma^{j'}({u_e})=v_e}+\II_{\tau^j({u_e})=\tau^{j'}({u_e})=v_e}\Bigg)\times\notag\\
    &\quad\prod_{j\in[\ell]}\left(\frac{1}{p}\right)^{\sum_{u\in \cA_j}\II_{\sigma^j(u)=\tau^j(u)}}\Bigg].\label{eq:h_mu_H_mu}
\end{align}

We thus get, using again Lemma \ref{lem:stein_chen_permutations}, Equation \eqref{eq:lemma_2_eq2} and noting that the conditioning on $\sigma^j({u_e})=v_e$ does not change the equivalent in \eqref{eq:lemma_2_eq2}:
\begin{align}
    H_\mu&\le\displaystyle (1+o(1))e^{[-1+1/p]\ell} (2/m)^{\ell}\times\notag\\
    &\qquad\qquad\dE\left[\prod_{\{j,j'\}\in \mu}
(a_{\{j,j'\}} b_{\{j,j'\}})\right]\\
    &=\displaystyle (1+o(1))e^{[-1+1/p]\ell} (2/m)^{\ell}\times\notag\\
    &\qquad\qquad\dE\left[\prod_{\{j,j'\}\in \mu}
(d_{\{j,j'\}}+e_{\{j,j'\}}) b_{\{j,j'\}}\right],
\end{align}
where $a_{\{j, j'\}}$, $b_{\{j, j'\}}$, $d_{\{j, j'\}}$, $e_{\{j, j'\}}$ follow the definitions in~\eqref{eq:ac}-~\eqref{eq:ec}.

Now with probability $1-o(1)$, $d_{\{j,j'\}}=\II_{i(j)=i(j')}$ (see formula \eqref{eq:D'}). Also, by Lemma \ref{lem:poisson_intersections} we may replace $e_{\{j,j'\}}$ and $b_{\{j,j'\}}$ by independent Poisson random variables with parameters $(1+o(1))m^2/n$, so that
\begin{align}
    H_\mu&\le\displaystyle (1+o(1))e^{[-1+1/p]\ell} (2/m)^{\ell}\times\notag\\
    &\qquad\qquad\prod_{\{j,j'\}\in \mu}[(\II_{i(j)=i(j')}+m^2/n)m^2/n]\\
    &=(1+o(1))\displaystyle e^{[-1+1/p]\ell}2^{\ell}n^{-\ell/2}\times\notag\\
    &\qquad\qquad\prod_{\{j,j'\}\in \mu}[\II_{i(j)=i(j')}+m^2/n].
\end{align}

\paragraph{\textbf{Step 3.b.b: Evaluate Lower Bound.}} 
We now proceed to obtain a matching lower bound. We have
\begin{align}
    h_\mu\le \dE\left[\II_{f_j=1\forall j}\II_\cN\II_\cM\II_{M=\mu} \prod_{j\in[\ell]}\left(\frac{1}{p}\right)^{\sum_{u\in \cA_j}\II_{\sigma^j(u)=\tau^j(u)}}\right],
\end{align}
with
\begin{align}
    h_\mu&=\displaystyle \dE\Bigg[\II_{f_j=1\forall j}\II_\cN\II_\cM\II_{M=\mu} \sum_{(u_e,v_e)\in \cA_j\cap\cA_{j'}\times \cB_j\cap \cB_{j'}, e=\{j,j'\}\in \mu}\notag\\
    &\displaystyle\prod_{e=\{j,j'\}\in\mu}\Bigg(\II_{\sigma^j({u_e})=\sigma^{j'}({u_e})=v_e}+\II_{\sigma^j({u_e})=\tau^{j'}({u_e})=v_e}\notag\\
    &\quad+\II_{\tau^j({u_e})=\sigma^{j'}({u_e})=v_e}+\II_{\tau^j({u_e})=\tau^{j'}({u_e})=v_e}\Bigg)\times\notag\\
    &\quad\prod_{j\in[\ell]}\left(\frac{1}{p}\right)^{\sum_{u\in \cA_j}\II_{\sigma^j(u)=\tau^j(u)}}\Bigg].\label{eq:h_mu}
\end{align}
The probability that $f_j=1$ conditionally on the presence of edge $(u_e,v_e=\sigma^j(u_e))$ in $F_j$ is lower-bounded by
\begin{align}
    [1-2b_j a_j/m]_+.
\end{align}
Indeed, the probability that there is another edge within $F_j$ is upper-bounded by the expected number $2 a_jb_j/m$ of such edges. 
Let event 
\begin{align}
    \mathcal{L} &:= \{f_j=1, \forall j\in[\ell]\; \&\; \notag\\
    &\qquad\sigma^j({u_e})=\sigma^{j'}({u_e})=v_e, \forall e=\{j,j'\}\in \mu\}
\end{align}
{
Using the fact that conditionally on the event $\mathcal{L}$ for some specific choice of edges $(u_e,v_e)\in \cA_j\cap \cA_{j'}\times \cB_j\cap\cB_{j'}$, the events $\cN$, $\cM$ and $M=\mu$ are necessarily satisfied, by symmetry, 
\begin{align}
    h_\mu&\ge\displaystyle
\dE\Bigg[ \left(\prod_{j\in [\ell]}[1-2b_j a_j/m]_+\right)\II_\cN\times  \notag\\
&\qquad\left(\prod_{e=\{j,j'\}\in \mu}[a_{\{j,j'\}}b_{\{j,j'\}}](2/m)^\ell\right)\times\notag\\
&\qquad\left(\prod_{j\in[\ell]}\left(\frac{1}{p}\right)^{\sum_{u\in \cA_j}\II_{\sigma^j(u)=\tau^j(u)}}\right)\Bigg|\mathcal{L}\Bigg],
\end{align}
where $a_{\{j, j'\}}$ and $b_{\{j, j'\}}$ follow the definitions in~\eqref{eq:ac} and \eqref{eq:bc} respectively.
}

Consider the set $\hat \cA_j= \cA_j \cap \cup_{j'\in\{0,\ldots,\ell\}\setminus\{j\}}\cA_j$,  of size $a_j$, and the subets $\tilde \cA_{j,\sigma}:=\sigma^j(\hat\cA_j)$, $\tilde\cA_{j,\tau}:=\tau^j(\hat\cA_j)$. Consider also $\hat\cB_j:=\cB_j \cap \cup_{j'\in\{0,\ldots,\ell\}\setminus\{j\}}\cB_j$, a set of size $b_j$, and $\tilde\cB_{j,\sigma}:=(\sigma^{j})^{-1}(\hat\cB_j)$, $\tilde\cB_{j,\tau}:=(\tau^{j})^{-1}(\hat\cB_j)$.

The conditional expectation
\begin{align}
    \dE\left[\left.\left(\frac{1}{p}\right)^{\sum_{u\in \cA_j}\II_{\sigma^j_u=\tau^j_u}}\right|f_j=1\; \&\; \sigma^j_{u_e}=v_e\right]
\end{align}
can be lower-bounded by $\dE[p^{-Z}]$ where $Z$ is the number of common edges induced by two independent random bijections $\sigma\in\hbox{Bij}(U,V)$, $\tau\in\hbox{Bij}(U',V')$, where 
\begin{align}
    |U|,|V|,|U'|,|V'|,|U\cap U'|, |V\cap V'|=m-O(a_j+b_j).
\end{align}

We may now use Lemma~\ref{lem:stein_chen_permutations}, \eqref{eq:lemma_2_eq2} to obtain that, since with high probability, $a_j,b_j=o(m)$,
\begin{align}
    \dE[p^{-Z}]\ge e^{-1+1/p}(1+o(1)).
\end{align}

This yields, noting 
\begin{align}
    X_{tot}:=\sum_{|\cC|\ge 2}e_\cC, \; Y_{tot}:=\sum_{|\cC|\ge 2}b_\cC,
\end{align}
the following:
\begin{align}
    h_{\mu}&\ge (1+o(1))(2/m)^{\ell}(e^{-1+1/p})^{\ell}\times\notag\\
    &\qquad \dE\left[\II_\cN\II_{X_{tot}\le n^{1/4}, Y_{tot}\le n^{1/4}}\prod_{e=\{j,j'\}\in \mu}[a_{\{j,j'\}}b_{\{j,j'\}}]\right].
\end{align}
As in the evaluation of $H_\mu$ we can now use formula \eqref{eq:D'} and Lemma \ref{lem:poisson_intersections} to obtain
\begin{align}
    h_\mu &\ge (1+o(1)) 2^\ell (e^{-1+1/p})^{\ell}n^{-\ell/2} \times\notag\\
    &\qquad \prod_{e=\{j,j'\}\in \mu}[\II_{i(j)=i(j')}+m^2/n].
\end{align}

\subsection{\textbf{Step 4: Combine all parts to conclude the proof.}}
By Steps 1-3, we have shown that the joint moment is dominated by $\psi_1$ and that terms $\phi_2$ and $\psi_2$ are negligible. 
Plugging the matching upper bound of $H_u$ and lower bound of $h_u$ back to~\eqref{eq:psi_1} gives the final asymptotic form. 
We have thus established Theorem~\ref{thm:moment}.
\end{proof}

\section{Proof of Proposition~\ref{prop:bias}}\label{sec:proof-prop-bias}

\subsection{Proof of~\eqref{eq:asymp_expected_weight}}

\begin{proof}
We evaluate the expectation  \eqref{eq:W_expectation},
\begin{align}
\mathbb{E}[W] &= \mathbb{E} \left[ p^{-S_1} \cdot p^{-S_2} \right], \\
S_1 &= \sum_{u \in \mathcal{A} \cap \mathcal{A}_0} \left( \mathbb{I}_{\sigma(u) \in \mathcal{B}_0} + \mathbb{I}_{\tau(u) \in \mathcal{B}_0} \right),\\
S_2 &= \sum_{u \in \mathcal{A}} \mathbb{I}_{\sigma(u) = \tau(u) \,\&\, (u, \sigma(u)) \notin \mathcal{A}_0 \times \mathcal{B}_0},
\end{align}
where we wrote $W$ instead of $W(i)$ for notational convenience. 
Let $S':=\sum_{u\in \cA}\II_{\sigma(u)=\tau(u)}$. Clearly, 
$S'\le S_1+S_2$, so that
\begin{align}
\dE[W]&\ge \dE\left[p^{-S'}\right]=e^{(1/p)-1}(1+o(1)),
\end{align}
by Lemma \ref{lem:stein_chen_permutations}, Equation \eqref{eq:lemma_2_eq2}.

To obtain a matching upper bound, introduce the notations $a_0:=|\cA\cap \cA_0|$ and $b_0:=|\cB\cap \cB_0|$. Write
\begin{align}\label{eq:upper_W_split}
\dE[W]=\dE\left[W\II_{a_0,b_0\le m^{1/3}}\right]+\dE\left[W\II_{a_0\vee b_0>m^{1/3}}\right].
\end{align}
We first upper-bound the last term by writing
\begin{align}
&\left(\dE\left[W\II_{a_0\vee b_0>m^{1/3}}\right]\right)^2 \notag\\
& \le \dE\left[p^{-2S_1}\II_{a_0\vee b_0>m^{1/3}}\right]\dE\left[p^{-2S_2}\II_{a_0\vee b_0>m^{1/3}}\right]\\
&\le \dE\left[p^{-2S_1}\II_{a_0\vee b_0>m^{1/3}}\right] \dE\left[p^{-2S_2}\right]\\
&\le \dE\left[p^{-2S_1}\II_{a_0\vee b_0>m^{1/3}}\right]e^{(1/p)-1}(1+o(1)),
\end{align}
where we used Cauchy-Schwarz inequality in the first step, and Lemma \ref{lem:stein_chen_permutations}, Equation \eqref{eq:lemma_2_eq2} in the last step. Now conditionally on $a_0,b_0$, $S_1$ is upper-bounded by $2 (a_0\wedge b_0)$. This yields
\begin{align}
\dE\left[p^{-2 S_1}\II_{a_0\vee b_0> m^{1/3}}\right]\le 2 \cdot \dE\left[p^{-4 a_0}\II_{a_0> m^{1/3}}\right].   
\end{align}
We can invoke Lemma \ref{lem:poisson_intersections} to bound the distribution of $a_0$ by that of a random variable $X\sim \text{Poi}(Km/n)$. This yields
\begin{align}
\dE\left[p^{-4 a_0}\II_{a_0> m^{1/3}}\right]\le \dE \left[p^{-4 X} \II_{X> m^{1/3}}  \right]=o(1).
\end{align}
Combined, the last three displays yield
\begin{align}
\dE\left[W\II_{a_0\vee b_0>m^{1/3}}\right]=o(1). 
\end{align}
We now proceed to upper bound the first term in \eqref{eq:upper_W_split}.
Let $X_\sigma$ (respectively, $X_\tau$) denote the set of elements $u\in \cA\cap\cA_0$ such that $\sigma(u)\in \cB_0$ (respectively, such that $\tau(u)\in \cB_0$). Let also $Y_\sigma:=\sigma(X_\sigma)$ and $Y_\tau:=\tau(X_\tau)$. Consider two independent uniform random permutations $\tilde \sigma$, $\tilde \tau$ from $\text{Bij}(\cA\setminus X_\sigma,\cB\setminus Y_\sigma)$ and $\text{Bij}(\cA\setminus X_\tau,\cB\setminus Y_\tau)$ respectively. Finally, introduce the events 
\begin{align}
E_\sigma&=\{\forall u\in \cA_0\setminus X_\sigma,\tilde\sigma(u)\notin \cB_0\},\\
E_\tau&=\{\forall u\in \cA_0\setminus X_\tau,\tilde\tau(u)\notin \cB_0\}.
\end{align} 
Introduce the notations
\begin{align}
S''&:=\sum_{u\in \cA\setminus X_\sigma}\II_{\tilde\sigma(u)=\tilde\tau(u)},
\end{align}
and 
\begin{align}
\cF_0:=(\cA,\cB,\cA_0,\cB_0,X_\sigma, X_\tau,Y_\sigma, Y_\tau).
\end{align}
One then has
\begin{align}
&\dE\left[W\II_{a_0,b_0\le m^{1/3}}\right]\notag\\
&=\dE\left[ p^{-S_1}\II_{a_0,b_0\le m^{1/3}}\dE\left[ p^{-S''}|E_\sigma,E_\tau,\cF_0\right]\right]\\
&\le \dE\left[p^{-S_1}\II_{a_0,b_0\le m^{1/3}} \frac{\dE\left(p^{-S''}|\cF_0\right)}{\dP(E_\sigma\cap E_\tau|\cF_0)}\right]\cdot
\end{align}
By Lemma \ref{lem:stein_chen_permutations}, Equation \eqref{eq:lemma_2_eq2}, one has
$$
\dE\left(p^{-S''}|\cF_0\right)=e^{(1/p)-1}(1+o(1)).
$$
Let again $a_0:=|\cA\cap \cA_0|$, $b_0:=|\cB\cap\cB_0|$. We then have
$$
\dP(E_\sigma|\cF_0)\ge 1-\frac{a_0 b_0}{m-b_0}.
$$
Indeed, the probability that $\tilde \sigma$ induces an edge precluded by the event $E_\sigma$ is upper-bounded by the expected number of such edges, which is no larger than $a_0b_0/(m-b_0)$. Thus on the event $\{a_0,b_0\le m^{1/3}\}$, the event $E_\sigma\cap E_\tau$ has conditional probability $1-o(1)$, so that
\begin{align}
\dE[W\II_{a_0,b_0\le m^{1/3}}]&\le e^{(1/p)-1}(1+o(1))\dE\left[p^{-S_1}\II_{a_0,b_0\le m^{1/3}}\right].
\end{align}

Now, the number $\sum_{u\in \cA \cap \cA_0}\II_{\sigma(u)\in\cB_0}$ conditioned on $\cA,\cA_0,\cB,\cB_0$ is stochastically less than  a binomial random variable with parameters $(b_0,a_0/(m-b_0))$, where $a_0:=|\cA_0\cap \cA|$, $b_0:=|\cB_0\cap \cB|$. Thus
\begin{align}
    &\dE\left[p^{-S_1}|\cA,\cA_0,\cB,\cB_0\right]\\
    &=\left(\dE\left[p^{-\sum_{u\in \cA\cap \cA_0}\II_{\sigma(u)\in\cB_0}}|\cA,\cA_0,\cB,\cB_0\right] \right)^2\\
    &\le\left[1-\frac{a_0}{m-b_0}+\frac{a_0}{m-b_0}(1/p)\right]^{2 b_0}\\
    &\le e^{2(1/p-1)a_0 b_0/(m-b_0)}, 
\end{align}
and this last term equals $1+o(1)$ on the event $\{a_0,b_0\le m^{1/3}\}$.  
This allows to conclude \eqref{eq:asymp_expected_weight}.
\end{proof}

\subsection{Proof of~\eqref{eq:asymp_bias}}
\begin{proof}

We first pick $\tilde{A}$ uniformly at random from $\binom{V\setminus\{i\}}{m-1}$. 
Let $\cA = \tilde{A} \cup \{i\}$.
Let $\cB \in \binom{V'}{m}$ chosen uniformly at random. 
Let $\sigma, \tau \in \hbox{Bij}(\mathcal{A}, \mathcal{B})$ independently chosen uniformly at random.
If $\tilde{A}$ contains $i'$, we let $\cA' = \cA$, $\sigma' = \sigma$, $\tau' = \tau$.
Then, we have a zero contribution to the bias.
If $\tilde{A}$ does no contains $i'$, which happens with probability $\frac{\binom{n-2}{m-1}}{\binom{n-1}{m-1}}$, we define sets and bijections:
\begin{itemize}
    \item \(\mathcal{A} := \{i\} \cup \tilde{\mathcal{A}}, \quad \mathcal{A}' := \{i'\} \cup \tilde{\mathcal{A}}\),
    \item \(\tilde{\mathcal{A}} \in \binom{V \setminus \{i, i'\}}{m - 1}\) is chosen uniformly at random,
    \item \(\mathcal{B} \in \binom{V'}{m}\) is chosen uniformly at random,
    \item \(\sigma, \tau \in \hbox{Bij}(\mathcal{A}, \mathcal{B})\) and \(\sigma', \tau' \in \hbox{Bij}(\mathcal{A}', \mathcal{B})\) are independent, with \(\sigma'|_{\tilde{\mathcal{A}}} = \sigma|_{\tilde{\mathcal{A}}}\) and \(\tau'|_{\tilde{\mathcal{A}}} = \tau|_{\tilde{\mathcal{A}}}\) and $\sigma'({i'})=\sigma(i)$, $\tau'({i'})=\tau(i)$.
\end{itemize}

Using the expressions~\eqref{eq:W_as_sum} and~\eqref{eq:W_expectation} and observing that the summands for \(W(i)\) and \(W(i')\) differ only in the vertex being included in the subset \(\mathcal{A}\) versus \(\mathcal{A}'\), we can express the bias \(\dE[W(i)] - \dE[W(i')]\) as
\begin{align}
    \dE[W(i)] - \dE[W(i')] 
    &= \frac{\binom{n-2}{m-1}}{\binom{n-1}{m-1}} \dE\left[ p^{-(s + t)} - p^{-(s' + t')} \right], \label{eq:expr_beta}
\end{align}
where 
the exponents are defined as follows:
\begin{align}
    s &:= \sum_{u \in \mathcal{A} \cap \mathcal{A}_0} \left( \II_{\sigma(u) \in \mathcal{B}_0} + \II_{\tau(u) \in \mathcal{B}_0} \right), \\
    t &:= \sum_{u \in \mathcal{A}} \II_{\sigma(u) = \tau(u) \& (u, \sigma(u)) \notin \mathcal{A}_0 \times \mathcal{B}_0}, \\
    s' &:= \sum_{u \in \mathcal{A}' \cap \mathcal{A}_0} \left( \II_{\sigma'(u) \in \mathcal{B}_0} + \II_{\tau'(u) \in \mathcal{B}_0} \right), \\
    t' &:= \sum_{u \in \mathcal{A}'} \II_{\sigma'(u) = \tau'(u) \& (u, \sigma'(u)) \notin \mathcal{A}_0 \times \mathcal{B}_0}.
\end{align}

Since \(m = \Theta(\sqrt{n})\), the prefactor satisfies
\begin{align}
    \frac{\binom{n-2}{m-1}}{\binom{n-1}{m-1}} = 1 + o(1).
\end{align}
Moreover, the contributions from the shared subset \(\tilde{\mathcal{A}}\) factorize, allowing us to write
\begin{align}
    &\dE[W(i)] - \dE[W(i')] \notag\\
    &= (1 + o(1)) \dE\left[\left(p^{-\delta(i)} - p^{-\delta(i')}\right) \cdot p^{-(\tilde{s} + \tilde{t})} \right],
\end{align}
where
{
\begin{align}
    \delta(i) &:= \II_{\sigma(i) \in \mathcal{B}_0} + \II_{\tau(i) \in \mathcal{B}_0} + \II_{\sigma(i) = \tau(i) \& \sigma(i) \notin \mathcal{B}_0}, \\
    \delta(i') &:= \II_{\sigma'(i') = \tau'(i')}, \\
    \tilde{s} &:= \sum_{u \in \tilde{\mathcal{A}} \cap \mathcal{A}_0} \left( \II_{\sigma(u) \in \mathcal{B}_0} + \II_{\tau(u) \in \mathcal{B}_0} \right), \\
    \tilde{t} &:= \sum_{u \in \tilde{\mathcal{A}}} \II_{\sigma(u) = \tau(u) \& (u, \sigma(u)) \notin \mathcal{A}_0 \times \mathcal{B}_0}.
\end{align}
Note that $\delta(i')$ is simplified using the fact that $i \in \cA_0$ and $i' \notin \cA_0$.

We shall use { the fact that $\sigma'({i'})=\sigma(i)$ and $\tau'({i'})=\tau(i)$.} 
Conditioning on all the sets $\tilde\cA,\cA_0,\cB_0$, the first part in parentheses, $p^{-\delta(i)} - p^{-\delta(i')}$, evaluates to
\begin{align}
\begin{cases}
    p^{-2} - p^{-1} & \text{if } \sigma(i) = \tau(i) \in \mathcal{B}_0, \quad \text{w.p. } \frac{b_0}{m^2}, \\[4pt]
    p^{-2} - 1 & \text{if } \sigma(i) \ne \tau(i),\; \sigma(i), \tau(i) \in \mathcal{B}_0, \quad \text{w.p. } \frac{b_0(b_0 - 1)}{m^2}, \\[4pt]
    p^{-1} - 1 & \text{if } \II_{\sigma(i)\in\cB_0}+\II_{\tau(i)\in\cB_0}=1, \quad \text{w.p. } 2\frac{b_0}{m}(1 -\frac{b_0}{m}).
\end{cases}
\end{align}

Also, the second factor can be evaluated to $(1+o(1))e^{(1/p)-1}$ as in the proof of \eqref{eq:asymp_expected_weight}. We thus have
\begin{align}
    &\dE [W(i)] - \dE [W(i')]\notag\\ 
    &= (1 + o(1)) e^{(1/p) - 1} \cdot \dE\Bigg[
        \frac{b_0}{m^2}(p^{-2} - p^{-1})+ \notag\\
    &\qquad\qquad \frac{b_0(b_0 - 1)}{m^2}(p^{-2} - 1) 
        + \frac{2b_0(m - b_0)}{m^2}(p^{-1} - 1)
    \Bigg].
\end{align}

By Lemma \ref{lem:poisson_intersections}, we can treat $b_0$ as a Poisson  random variable with parameter $Km/n$.} The first two terms in brackets in the last display thus have expectation of order $O(K/(mn))$, while the last term has expectation of order $K/n$, which is larger. We thus have
\begin{align}
    \dE [W(i)] -\dE [W({i'})]=(1+o(1))e^{(1/p)-1}[(1/p)-1]\frac{2K}{n},
\end{align}
as announced.
\end{proof}

\section{Proof of Corollary~\ref{coro:GBS-usefulness}}\label{sec:proof-coro-GBS-usefulness}

\begin{proof}
    Let $\Phi_0$ and $\Phi_1$ be the CDF of $f_0$ and $f_1$ respectively. 
    Let $h_{cn}$ be the threshold, above which are the top $cn$ samples out of all $n$ samples.
    We approximate the threshold $h_{cn}$ by the top $c$-quantile of $f_0$, i.e.,
    \begin{align}
        h_{cn} \approx \Phi_0^{-1}(1-c).
    \end{align}
    Then, the probability that a sample from $f_1$ exceeds $h_{cn}$ is approximately
    \begin{align}
        1- \Phi_1(\Phi_0^{-1}(1-c)).
    \end{align}
    Taking $\epsilon\sqrt{n}$ samples from $f_1$, the expected number of samples whose value exceeds $h_{cn}$ is approximately
    \begin{align}
        &\epsilon\sqrt{n}(1- \Phi_1(\Phi_0^{-1}(1-c)))\notag\\
        &=\epsilon\sqrt{n}\left(1- \Phi_1\left(\mu + \frac{\mu}{\sqrt{n}}\sqrt{2}\cdot\hbox{erf}^{-1}(1-2c)\right)\right)\\
        &=\epsilon\sqrt{n}\left(1- \tilde{\Phi}\left(\frac{\mu + \frac{\mu}{\sqrt{n}}\sqrt{2}\cdot\hbox{erf}^{-1}(1-2c)-\mu + \frac{\epsilon\sqrt{n}}{n}\mu}{\frac{\mu}{\sqrt{n}}}\right)\right)\\
        &=\epsilon\sqrt{n}\left(1- \tilde{\Phi}\left(\sqrt{2}\cdot\hbox{erf}^{-1}(1-2c)-\epsilon\right)\right)\\
        &=\epsilon\sqrt{n}\left(1- \tilde{\Phi}\left(\tilde{\Phi}^{-1}(1-c)-\epsilon\right)\right). \label{eq:proportion-1}
    \end{align}
    If $\epsilon \rightarrow 0$, then~\eqref{eq:proportion-1} converges to $c\epsilon\sqrt{n} \ll  cn$. As $\epsilon$ increases, the mean shift of $f_1$ makes samples from $f_1$ increasingly likely to be in the top ranks.
\end{proof}

\section{Proof of Theorem~\ref{thm:lognormal}}\label{sec:proof-thm-lognormal}

\begin{proof}
We begin by establishing the following two lemmas, which together directly imply Theorem~\ref{thm:lognormal}.

Let 
\begin{align}
    \hat{\text{Haf}}(M) := \prod_{i=1}^n\frac{\sum_{j=1}^n\xi_{ij}}{n p}.\label{eq:hat-haf-M}
\end{align}
    
\begin{lemma}\label{lem:1}
For fixed $p\in (0,1)$, the random variable $\hat{\text{Haf}}(M)$ converges weakly, as $n\to \infty$, to the log-normal distribution with parameters $-\frac{1-p}{2p}$ and $\frac{1-p}{p}$. In other words it converges weakly to $\exp(X)$ where $X\sim \cN(-\frac{1-p}{2p}, \frac{1-p}{p})$.
\end{lemma}
\begin{proof}
Write
\begin{align}
    \hat{\text{Haf}}(M) = \exp\left( \sum_{i=1}^n\ln\left(1+\frac{\sqrt{1-p}}{\sqrt{np}}U_i\right)\right),
\end{align}
where we introduced the notation 
\begin{align}
    U_i :=\frac{1}{\sqrt{(1-p)np}}\sum_{j=1}^n(\xi_{ij}-p).
\end{align}
Thus, taking the logarithm in the previous identity and taking a Taylor expansion to the second order of the logarithm, we get
\begin{align}\label{eq:log_x_hat}
\ln(\hat{\text{Haf}}(M))=\sum_{i=1}^n \frac{\sqrt{1-p}}{\sqrt{np}}U_i -\frac{1}{2}\sum_{i=1}^n \frac{1-p}{np} U_i^2 +R_n,
\end{align}
where 
\begin{align}
    R_n:=\sum_{i=1}^n \ln\left(1+\frac{\sqrt{1-p}}{\sqrt{np}}U_i\right)-\frac{\sqrt{1-p}}{\sqrt{np}}U_i+\frac{1}{2}\frac{1-p}{np} U_i^2.
\end{align}
Using Chernoff-like bounds, we have that 
\begin{align}
    \dP\left(|U_i|\ge A\right)\le 2 \exp(-C A^2),
\end{align}
for all $A\in [0, n^{1/2-\epsilon}]>0$, for any fixed $\epsilon>0$, and some suitable constant $C>0$. 
By choosing $A=\alpha\sqrt{\ln(n)}$ and taking a union bound, we obtain
\begin{align}
    \dP\left(\exists i\in[n]:|U_i|\ge \alpha \sqrt{\ln(n)}\right)\le 2n^{1-C \alpha^2},
\end{align}
which is $o(1)$ for $\alpha>1/\sqrt{C}$, which we assume to hold.
On the event $\left\{\forall i\in[n],\; |U_i|\le \alpha \sqrt{\ln(n)}\right\}$, which has probability $1-o(1)$, it holds that 
\begin{align}
    |R_n| &\le n \sup_{|x|\le \alpha\sqrt{\frac{(1-p)\ln(n)}{np}}}\left|\ln(1+x)-x +\frac{1}{2}x^2\right|\\
    &\le n \left( \alpha\sqrt{\frac{(1-p)\ln(n)}{np}}\right)^3\\
    &=O\left(n^{-1/2}(\ln n)^{3/2}\right)=o(1).
\end{align}

By the central limit theorem, $\sum_{i=1}^n \frac{\sqrt{1-p}}{\sqrt{pn}} U_i$ converges weakly as $n\to\infty$ to $\cN(0,\frac{1-p}{p})$. 
By the weak law of large numbers (which holds in the present case as can be seen from applying Bienaymé-Chebyshev inequality), $\sum_{i=1}^n \frac{1-p}{pn} U_i^2$ converges in probability to $\frac{1-p}{p}$. 
Together, these properties and \eqref{eq:log_x_hat} imply that $\ln(\hat{\text{Haf}}(M))$ converges weakly as $n\to\infty$ to $\cN(-\frac{1-p}{2p}, \frac{1-p}{p})$. 
\end{proof}
We next need the following:
\begin{lemma}\label{lem:2}
For $\text{Haf}(M)$ and $\hat{\text{Haf}}(M)$ as defined in~\eqref{eq:haf-M} and~\eqref{eq:hat-haf-M}, one has
\begin{align}\label{eq:lim_zero}
\lim_{n\to\infty} \dE [(\text{Haf}(M)-\hat{\text{Haf}}(M))^2]=0
\end{align}
\end{lemma}
\begin{proof}
To establish \eqref{eq:lim_zero}, we expand the square and evaluate separately the limits of $\dE[\text{Haf}(M)^2]$, $\dE[\text{Haf}(M) \hat{\text{Haf}}(M)]$, and $\dE[\hat{\text{Haf}}(M)^2]$. Let us start with the latter.
Write
\begin{align}
    \dE[\hat{\text{Haf}}(M)^2] &= \dE \left[\prod_{i=1}^n \frac{1}{n^2 p^2}\left(\sum_{j=1}^n \xi_{ij}\right)^2\right]\\
    &=\left(\frac{1}{n^2 p^2}(n^2p^2+n(p-p^2))\right)^n\\
    &= \left(1+\frac{1-p}{np}\right)^n \rightarrow e^{(1-p)/p} \text{ as } n \rightarrow \infty.
\end{align}
Write next
\begin{align}
    \dE[\text{Haf}(M) \hat{\text{Haf}}(M)] &= \frac{1}{ p^{2n} n! n^n} \sum_{\sigma, \tau \in P^2_n} \prod_{i=1}^n \dE[\xi_{i\sigma(i)} \xi_{i\tau(i)}]\\
    &=\frac{1}{n! n^n} \sum_{\sigma, \tau \in P^2_n} \left(\frac{1}{p}\right)^{\sum_{i=1}^n \II_{\sigma(i) = \tau(i)}}\\
    &\overset{(a)}{=}\frac{1}{n^n} \sum_{\tau\in P^2_n}\left(\frac{1}{p}\right)^{\sum_{i=1}^n \II_{i = \tau(i)}}\\
    &\rightarrow e^{(1-p)/p}\hbox{ as }n\to\infty,
\end{align}
where in equality (a) we used the fact that, by symmetry, each permutation $\sigma$ contributes the same to the previous sum. 

Finally, we evaluate
\begin{align}
    \dE[\text{Haf}(M)^2] &= \frac{1}{p^{2n}(n!)^2} \sum_{\sigma, \tau \in P^2_n} \prod_{i=1}^n \dE[\xi_{i\sigma(i)} \xi_{i\tau(i)}]\\
    &= \frac{1}{(n!)^2} \sum_{\sigma, \tau \in P^2_n} \left(\frac{1}{p}\right)^{\sum_{i=1}^n \II_{\sigma(i) = \tau(i)}}\\
    &=\frac{1}{n!} \sum_{\tau\in P^2_n} \left(\frac{1}{p}\right)^{\sum_{i=1}^n \II_{i = \tau(i)}}.
\end{align}
This final expression reads
\begin{align}
    \dE[\text{Haf}(M)^2] =\frac{1}{n!}\sum_{i=0}^n \binom{n}{i} p^{-i} De(n-i),
\end{align}
where $De(j)$ denotes the number of permutations of $j$ elements having no fixed points. Using $De(j)\approx j! e^{-1} $, and the fact that $De(j)\le j!$, by writing 
\begin{align}
    \dE[\text{Haf}(M)^2] =\sum_{i=0}^n \frac{1}{i!} p^{-i}\frac{De(n-i)}{(n-i)!},
\end{align}
we can use Lebesgue's dominated convergence theorem to conclude that $\dE[\text{Haf}(M)^2]\to e^{1/p-1}$ as $n\to\infty$. We thus have that $\dE[\text{Haf}(M)^2]$, $\dE[\text{Haf}(M) \hat{\text{Haf}}(M)]$, and $\dE[\hat{\text{Haf}}(M)^2]$ all converge to $e^{(1-p)/p}$ as $n\to\infty$, which establishes the Lemma's claim. 
\end{proof}
Put together Lemma~\ref{lem:1} and Lemma~\ref{lem:2} readily imply Theorem~\ref{thm:lognormal}.
\end{proof}


\bibliography{ref}

\end{document}